\documentclass[12pt]{iopart}
	\usepackage{amssymb}
	\usepackage{epsfig,floatflt}
	\usepackage{exscale}
    \usepackage[normalem]{ulem}	

\begin{document}
\title[Positronium in a liquid phase]{Positronium in a liquid phase: formation, bubble state and chemical reactions}
\author{Sergey V.~Stepanov$^1$, Vsevolod M.~Byakov$^{1,2}$, Dmitrii S.~ Zvezhinskiy$^1$,
        Gilles Dupl\^atre$^3$, Roman R. Nurmukhametov$^1$ and Petr S. Stepanov$^1$}
\address{$^1$ Institute of Theoretical and Experimental Physics,
		Moscow 117218, Russia}
\address{$^2$ D.Mendeleyev University of Chemical Technology, Miusskaya sq., 9, Moscow 125047, Russia}
\address{$^3$ Institut Pluridisciplinaire Hubert Curien, \\ CNRS/IN2P3, BP 28 67037 Strasbourg, Cedex 2, France}
		\ead{stepanov@itep.ru}


\begin{abstract}
This chapter reviews the following items:

1. Energy deposition and track structure of fast positrons: ionization slowing down, number of ion-electron pairs, typical sizes, thermalization, electrostatic interaction between e$^+$ and its blob, effect of local heating;

2. Positronium formation in condensed media: the Ore model, quasifree Ps state, intratrack mechanism of Ps formation;

3. Fast intratrack diffusion-controlled reactions: Ps oxidation and ortho-para conversion by radiolytic products, reaction rate constants, interpretation of the PAL spectra in water at different temperatures;

4. Ps bubble models. ``Non-point'' positronium: wave function, energy contributions, relationship between the pick-off annihilation rate and the bubble radius.
\end{abstract}
\maketitle

Positrons (e$^+$) as well as positronium atoms (Ps) are recognized as nanoscale probes of the local structure in a condensed phase (liquid or solid) and of the early radiolytic physicochemical processes occurring therein. The parameters of positron annihilation spectra determined experimentally (e.g.,  positron and Ps lifetimes, angular and energetic widths of the spectra, Ps formation probability) are highly sensitive to the chemical composition, the local molecular environment of Ps (free volume size), and the presence of structural defects. They are also sensitive to variation of temperature, pressure, external electric and magnetic fields, phase transitions.

The informative potentiality of positron spectroscopy strongly depends on the reliability of any theory describing the behavior of positrons in matter, since it should help to decipher the information coded in the annihilation spectra. So, realistic (quantum mechanical) models are needed for e$^+$ track structure, e$^+$ energy losses, ionization slowing down and thermalization, intratrack reactions (ion-electron recombination, solvation, interaction with scavengers), Ps formation process, Ps interaction with chemically active radiolytic species and e$^+$/Ps trapping by structural defects.

Usually, treatment of the measured annihilation spectra is reduced to their resolution into a set of simple trial functions: sums of decaying time exponentials in the case of PALS (Positron Annihilation Lifetime Spectroscopy), of Gaussians in the case of ACAR (Angular Correlation of Annihilation Radiation) and DBAR (Doppler Broadening of Annihilation Radiation). The outcome of such conventional analyses of positron annihilation data are the intensities of these components and the corresponding lifetimes/widths.

However, realistic theoretical models suggest more complex kinetics for describing the intratrack reactions between the positron and Ps atom and radiolytic products and scavengers. For example, because of the inhomogeneous spatial distribution of the track species, their outdiffusion becomes an important factor: diffusion kinetics cannot be expressed in terms of mere exponentials or Gaussians. Obviously, elaborated theoretical models should be used in the fitting procedure of the annihilation spectra. In this case, the adjustable/fitting parameters (reaction rate constants or reaction radii, diffusion coefficients, initial size of the terminal part of the e$^+$ track, contact density in the Ps atom) would present a clear physical meaning instead of the above mentioned ``intensities'' of some trial functions.

The present chapter is an attempt to a systematic presentation of the theoretical model, which describes e$^+$ fate since its injection into a liquid until its annihilation. It should be a basis for a positron spectroscopy of liquid media.

\section{Energy deposition and track structure of the fast positron}

\subsection{Ionization slowing down}

Positrons, produced in nuclear $\beta^+$-decay, have initial energies of about several hundreds of keV. Once injected into a medium they lose energy via molecular ionization.  Within 10 ps the positron energy drops down to the ionization threshold.  Further approach to thermal equilibrium proceeds primarily via excitations of intra- and intermolecular vibrations. This usually takes several tens of picoseconds \cite{Dry11,Par87}.

\begin{figure}[!t]
\begin{center}
\epsfig{file=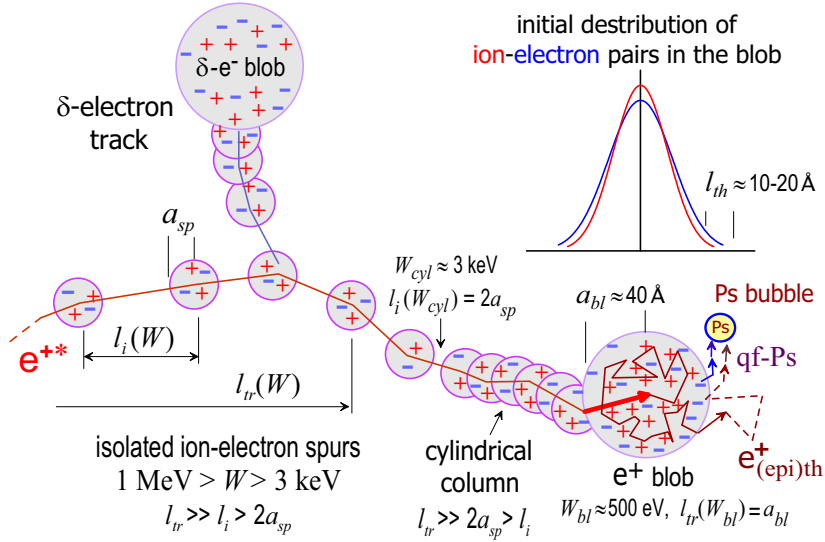, width=110mm}
\caption{Scheme of the end part of the e$^{+*}$ track and Ps
formation.} \label{e_track}
\end{center}
\end{figure}

Roughly half of the positron kinetic energy is lost in rare head-on collisions, resulting in the knocking out of $\delta$-electrons with kinetic energies of about several keV, the tracks of these electrons forming branches around the positron trail (Fig.~\ref{e_track}).  The other half of the energy is spent in numerous glancing collisions with molecules. The average energy loss in such a collision is 30-50 eV (at maximum,100 eV).  A secondary electron knocked out in a glancing collision produces, in turn, a few ion-electron pairs inside a spherical microvolume, called a ``spur'' in radiation chemistry.  Its radius, $a_{sp}$, is determined by the thermalization length of the knocked-out electrons in the presence of the Coulombic attraction of the parent ions. Note: why restrict to the electrons knocked-out "in a glancing collision"?

While the positron energy $W$ is greater than $W_{cyl}\sim 3$ keV, the mean distance $l_i$ between adjacent ionizations produced by the positron is greater than the spur size $2a_{sp}$. It ensues that at high positron energies, the spurs are well separated from each other. The trajectory of the fast positron is a quasi-straight line because $l_i$ is less than the positron transport path $l_{tr}$. The latter is the mean distance traveled by the positron before it changes its initial direction of motion by 90$^\circ$.

When $l_i<2a_{sp}<l_{tr}$ or $W_{bl}<W<W_{cyl}$ the spurs overlap, forming something like a cylindrical ionization column. When the e$^+$ energy becomes less than the blob formation energy, $W_{bl}$ ($\sim 500$ eV, see below), the positron is about to create a terminal blob. The diffusion motion of e$^+$ in the blob becomes more pronounced: the direction of its momentum changes frequently due to elastic scattering and the ionization of surrounding molecules. Roughly speaking, all intrablob ionizations are confined within a sphere of radius $a_{bl}$. The terminal positron blob contains a few tens of ion--electron pairs ($n_0 \approx W_{bl}/W_{iep} \approx 30$) because the average energy $W_{iep}$ required to produce one ion--electron pair is 16--22 eV \cite{Bya85}. Finally, the positron becomes subionizing and its energy loss rate drops by almost 2 orders of magnitude \cite{Ste95}.

\begin{figure}[!t]
\begin{minipage}[h]{75mm}
\epsfig{file=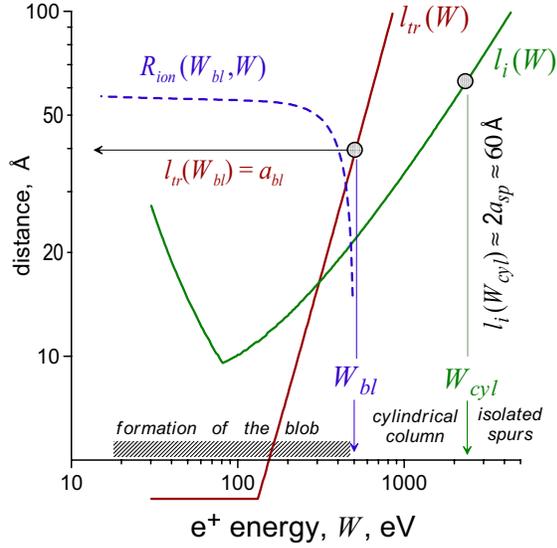, width=75mm}
\end{minipage}
\begin{minipage}[h]{75mm}
\caption{Dependences of $l_i(W)$ and $l_{tr}(W)$ vs the energy $W$ of the positron in liquid water.  This figure illustrates the solution of equations for the blob parameters $W_{bl}$ and $a_{bl}$: $l_{tr}(W_{bl}) = a_{bl}$, $2a^2_{bl} = R^2_{ion}(W_{bl},Ry)$.}\label{a_bl_W_bl}
\end{minipage}
\end{figure}

Typical dependencies of $l_i(W)$ and $l_{tr}(W)$ versus the energy of the positron are shown in Fig.~\ref{a_bl_W_bl}. The calculation of $l_i(W)=W_{iep}/{\rm LET}(W)$ is based on LET (Linear Energy Transfer) data of e$^\pm$.  The estimation of the transport path has been done in the framework of the Born approximation (the wavelength of the positron with energy $\gtrsim 100$ eV is small in comparison with the size of molecules), where the Born amplitude was calculated by simulating a molecule of the liquid as an iso-electronic atom.

At low energies, positron scattering becomes more and more efficient and its motion must be regarded as diffusion-like with the energy-dependent mean free path, $l_{tr}(W)$, between successive ``collisions'', which completely randomize the direction of the velocity of the particle. During ionization slowing down, when e$^{+*}$ loses energy from $W_i$ down to $W_f$ the mean square displacement of the positron is
\begin{equation}   \label{Eq1}
R^2_{ion}(W_i,W_f) = \int 6D_p(W) dt =
    \int_{W_f}^{W_i} l_{tr}(W) \frac{dW}{|-dW/dx|_{ion}}.
\end{equation}
where $D_p(W)= l_{tr}(W) v_p/6$ is the energy-dependent diffusion coefficient and $dt = \frac{dx}{v_p} = \frac{1}{v_p}\cdot \frac{dW}{|-dW/dx|_{ion}}$. Now we can define $W_{bl}$ and $a_{bl}$. These quantities are obtained from the following equations:
\begin{equation}   \label{Eq2}
l_{tr}(W_{bl}) = a_{bl},
    \qquad
a^2_{bl} = R^2_{ion}(W_{bl},{\rm Ry}) - a^2_{bl}.
\end{equation}
Here Ry=13.6 eV stands for a typical ionization potential. Eq.~(\ref{Eq2}) indicates that the terminal positron blob is a spherical nanovolume, which confines the end part of its trajectory.  This is where ionization slowing down is most efficient (the thermalization stage of the subionizing positron is not included here).  The mathematical formulation of this statement is twofold.  Just after the first ``step'' of the blob formation, which is equal to $l_{tr}(W_{bl})$ (the thick red arrow in Fig.~\ref{e_track}), the positron reaches the center of the blob. After that, the end part of the ionization slowing-down trajectory is embraced by the blob; i.e., the mean-square slowing-down displacement of the positron, $R^2_{ion}(W_{bl},{\rm Ry}) - a^2_{bl}$ is equal to $a^2_{bl}$ the ``radius'' of the blob squared.

The solution of these equations in the case of liquid water is shown in Fig.~\ref{a_bl_W_bl}.  It gives $W_{bl}\approx 500$ eV and $a_{bl}\approx 40$ \AA. One may assume that the values of $a_{bl}$ and $W_{bl}$ do not differ significantly from one liquid to another because the ionization slowing-down parameters depend mostly on the ionization potential and the average electron density, parameters that are more or less the same in all molecular media.  At first sight it seems strange that the obtained blob size $a_{bl}\approx 40$ \AA\ is smaller than the dimensions of ion-electron pairs (60-200 \AA), well known from radiation chemistry \cite{HRC91}.  This is because the dimension of the ion-electron pair is determined by the thermalization length of the subionizing electron (in the field of the parent cation), while the positron blob radius is determined by the ionization slowing-down of the energetic positron when it loses the last $\sim 500$ eV of its energy.

\subsection{Thermalization stage. Interaction between the positron and its blob}

At the end of the slowing-down by ionization and electronic excitation, the spatial distribution of e$^{+*}$ coincides with the distribution of the blob species (i.e., $\sim \exp (-r^2/a_{bl}^2)$).  Such a subionizing positron having some eV of excess kinetic energy may easily escape from its blob because the interaction between the blob and the e$^{+*}$ is rather weak (see next section).  It is expected that by the end of thermalization, the e$^+$ distribution becomes broader with the dispersion:
\begin{equation}   \label{Eq3}
a_p^2 \approx a_{bl}^2 + \left\langle R^2_{vib}(W_0,T) \right\rangle_{W_0}.
\end{equation}
Here $R^2_{vib}(T,W_0)$ is determined by Eq.~(\ref{Eq1}), where $|dW/dx|_{ion}$ should be replaced by $|dW/dx|_{vib}$,
i.e., the stopping power of subionizing e$^{+*}$ related to the excitation of vibrations ($T$ is the temperature in energy units). The estimation of $a_p$ requires quantitative data on $|dW/dx|_{vib}$, scattering properties of subionizing e$^{+*}$, and the spectrum of its initial energies $W_0$ after the last ionization event.  In Eq.~(\ref{Eq3}) $\langle \dots \rangle_{W_0}$ denotes the average over $W_0$. In contrast to the parameters related to ionization slowing down, $a_p$ strongly depends on the properties of each particular liquid and may reach hundreds of \AA\ especially in non-polar media.

Between the positively charged ions and the knocked-out intrablob electrons there exists a strong Coulombic attraction: out-diffusion of the electrons (even during their thermalization) is almost completely suppressed and the distribution of the ions is close enough to that of electrons. This case is known as ambipolar diffusion when ions and electrons expand with the same diffusion coefficient equal to the duplicated diffusion coefficient of the ions. Thus, the distribution of the blob electrons remains slightly broader than that of ions. So, the calculated electrostatic potential therein is everywhere repulsive towards the positron and its typical value is about several $T$.

However, there is an opposite effect: while residing inside the blob, the thermalized e$^+$ rearranges the intrablob electrons, so that the total energy of the system decreases because of the Debye screening.  The corresponding energy drop may be estimated using the Debye-Huckel theory. It is $\sim \frac{e^2}{\varepsilon (r_D + a_{bl}/n_0^{1/3})}$. However, the Debye radius $r_D \approx (4\pi r_c c_{iep})^{-1/2}\approx 4$ \AA\, is quite small in comparison with $a_{bl}/n_0^{1/3}$ the average distance between intrablob electrons (here $r_c = e^2/\varepsilon_\infty T$ is the Onsager radius and $c_{iep} \approx n_0 \left/ \frac{4}{3}\pi a_{bl}^3 \right.$ is the concentration of ion-electron pairs within the blob). Thus, this screening energy $\sim T n_0^{1/3} r_c/a_{bl}$ becomes a dominant contribution, which is some tenths of eV. It may result into e$^+$ trapping within the blob.

The above estimation shows that both the thermalized and epithermal positrons could be trapped within the blob. This effect is important to calculate the Ps formation probability in an external electric field \cite{Ste01MSF,Ste05PRB}.

\subsection{Effects of local heating and premelting in the terminal part of the e$^+$ track}

We have estimated above that the ionization slowing down of energetic positrons (e$^{+*}$) and subsequent ion-electron recombination release an energy up to 1 keV in the terminal e$^+$ blob, this energy being finally converted into heat. Therefore the temperature in the e$^+$ blob should be higher than the bulk temperature. This phenomenon may be called as {\it the local heating effect}. This transient temperature regime may strongly affect, for example, the Ps bubble growth, by changing viscosity of the medium\cite{09Ste_MSF_Ps_bub}. This effect may also influence the mobility of intratrack species and their reaction rates constants.

For quantitative estimations we simulate this process with the help of the macroscopic heat transfer equation:
\begin{equation}\label{LH1}
c_p \rho \frac{\partial T(r, t)}{\partial t} = \hbox{div}(\lambda \nabla T)  + q_+(r,t),
\qquad T(r,t=0) = T_{bulk}.
\end{equation}
Here $T(r,t)$ is the local temperature, $T_{bulk}$ is the bulk temperature of the medium, $c_p$ is its specific heat capacity, $\rho$ is the density, $\lambda$ is the thermal conductivity. The second term in the RHS quantifies the energy released by the positron when creating its blob: $q_+(r,t) \approx W_{bl} G(r,a) f(t,\tau)$, where $W_{bl}\approx 1$ keV is the blob formation energy, $G(r,a) \approx \frac{e^{-r^2/a_{bl}^2}}{\pi^{3/2}a_{bl}^3}$ describes the spatial distribution of the released energy and $f(t,\tau) \approx \exp \left( - \frac{(t-1 \rm{ps})^2}{2 \tau^2} \right) /(\sqrt{2 \pi}\tau )$ is its temporal distribution, where $\tau\approx 0.3$ ps is the typical time of ion-electron recombination.

\begin{figure}[!t]
\begin{center}
\epsfig{file=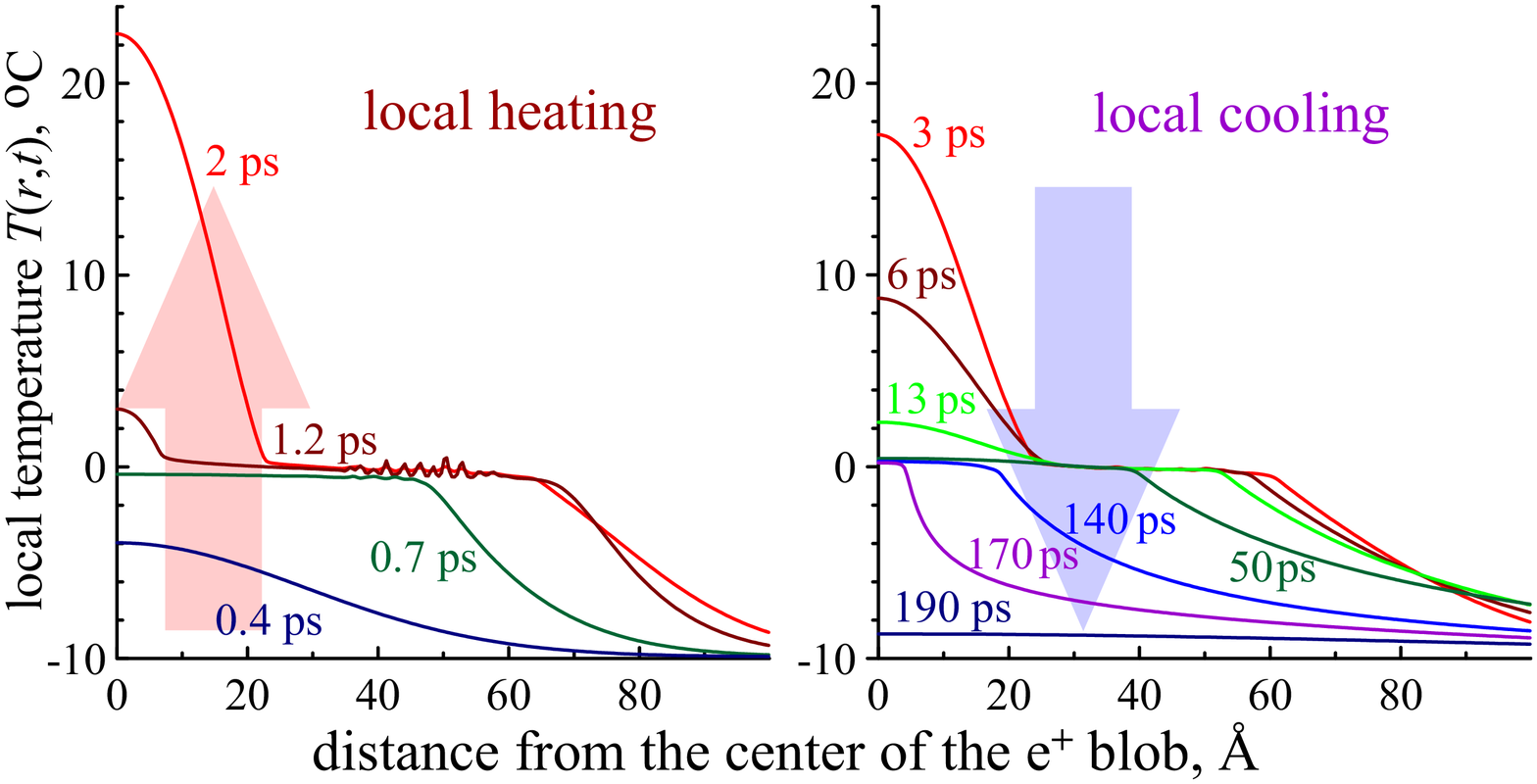, width=100mm}
\epsfig{file=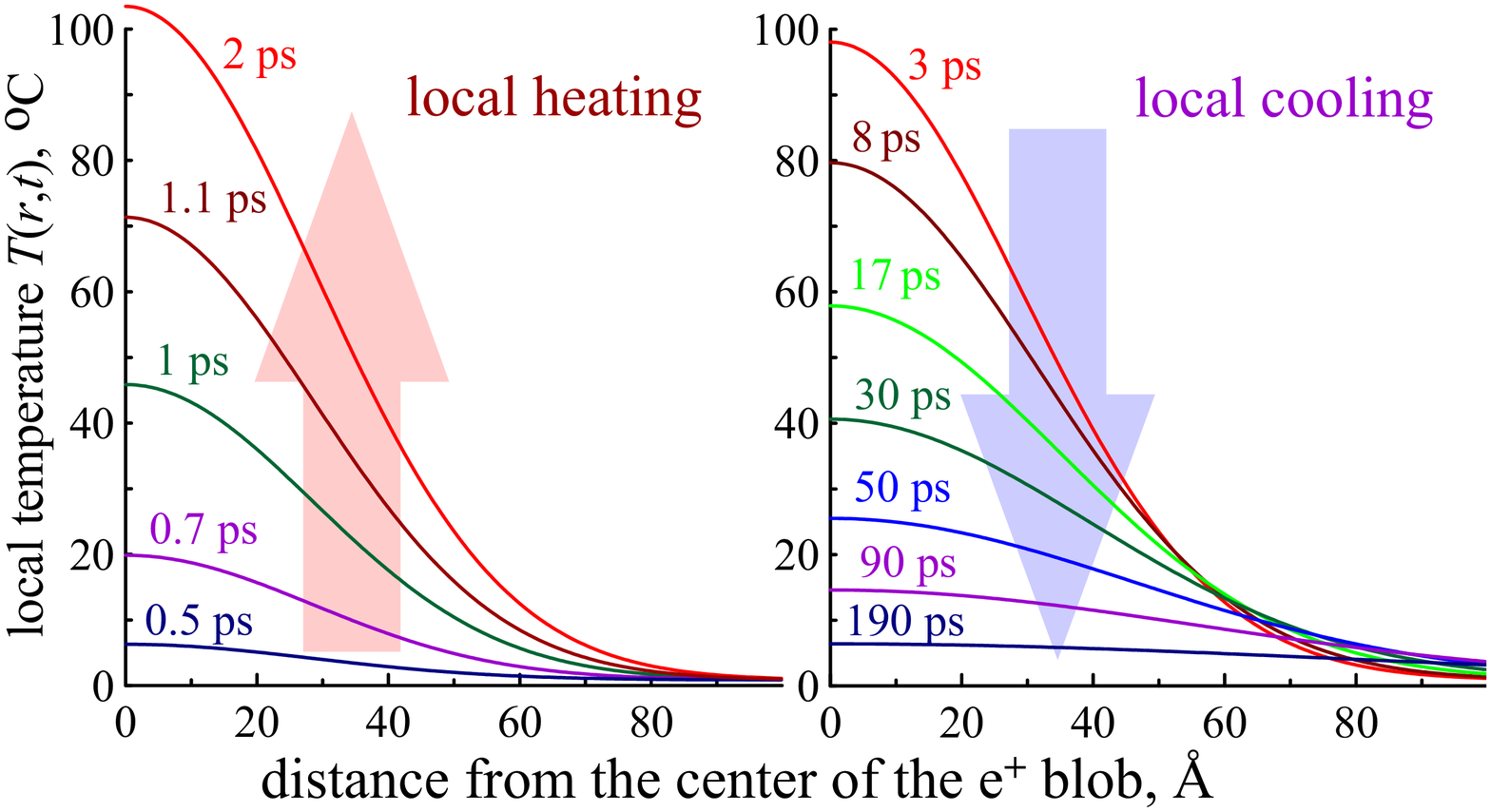, width=100mm}
 \caption{Calculated temperature profiles in ice $T_{bulk}=-10$ $^\circ$C (on top) and in water $T_{bulk}$ (bottom) is slightly above 0 $^\circ$C. $a_{bl}=40$ \AA, $W_{bl}=1$ keV.}
\label{loc_heating}
\end{center}
\end{figure}

\begin{table}[b]
\caption{Thermodynamical properties for water and some alcohols in solid and liquid phases near melting temperature \cite{cp_alcohols,Yaws,Kor11,dean, crc}} \label{TabProp}
\begin{center}
\begin{tabular}{|c|c|c|c|c|}
\hline
substance& $T_{m}$, K & $\lambda^S$ / $\lambda^L$; W/m/K  \cite{Kor11,dean}
		   & $q_m$; J/g \cite{crc}& $\rho^S$ / $\rho^L$; g/cm$^3$ \cite{Yaws}\\
\hline
methanol & 175.6 & 0.32/0.21 &  99 & 0.98/0.79\\
ethanol  & 159   & 0.27/0.17 & 109 & 1.06/0.79\\
butanol  & 184.5 & 0.47/0.16 & 125 & 1.05/0.81\\
water    & 273   & 2.38/0.56 & 334 & 0.92/1.00\\
\hline
\end{tabular}
\end{center}
\end{table}

In general, the system can be simultaneously solid and liquid. Then, the following method for obtaining a numerical solution of Eq. (\ref{LH1}) can be used \cite{Muh09,Bon73}. To describe the deposition of the latent heat of melting, $q_m$, we added to $c_p(T)$ a term, which is non-zero only in a narrow temperature interval $T_m-\Delta T<T<T_m$, where $T_m$ is the melting point temperature and $\Delta T$ is the width of the phase transition (arbitrarily fixed to 0.25 K). Moreover, an integration of this term over temperature must give $q_m$. This additional overshot to $c_p(T)$ is simulated by a Gaussian function as follows:
\begin{equation}\label{LH3}
\tilde c_p(T) = c_p(T) + q_m \exp\left(-\frac{(T-T_m)^2}{2 \Delta T^2}\right)\left/ \sqrt{2 \pi \Delta T^2}\right. .
\end{equation}
Close to the phase transition region, the $T$-dependencies of thermal conductivity and density were approximated by smooth functions:
\begin{equation}\label{LHLambda}
 \lambda(T)=\frac{\lambda^{S}}{\exp(-(T_{m}-T)/\Delta T)+1} + \frac{\lambda^{L}}{\exp( (T_{m}-T)/\Delta T)+1},
\end{equation}
$$
\rho(T)=\frac{\rho^{S}}{\exp(-(T_{m}-T)/\Delta T)+1} + \frac{\rho^{L}}{\exp( (T_{m}-T)/\Delta T)+1}.
$$
This approach works well especially when it is hard to trace an interphase boundary \cite{Muh09}.

Simulations were done for water, methanol, ethanol and butanol. Some thermodynamic properties of these substances are given in Table \ref{TabProp}.

Temperature profiles $T(r,t)$ were calculated numerically for $0<r<200$ \AA\ and $0<t<1$ ns. Eq. (\ref{LH1}) was solved as a 1-D problem with boundary conditions $T(r=200{\rm ~\AA}) =T_{bulk}$ with the help of \texttt{PDEPE} subroutine from \texttt{Matlab}. Some temperature profiles in ice and in water close to the melting point are shown in Fig. \ref{loc_heating}.

\begin{figure}
 \centering
\includegraphics[scale=0.5]{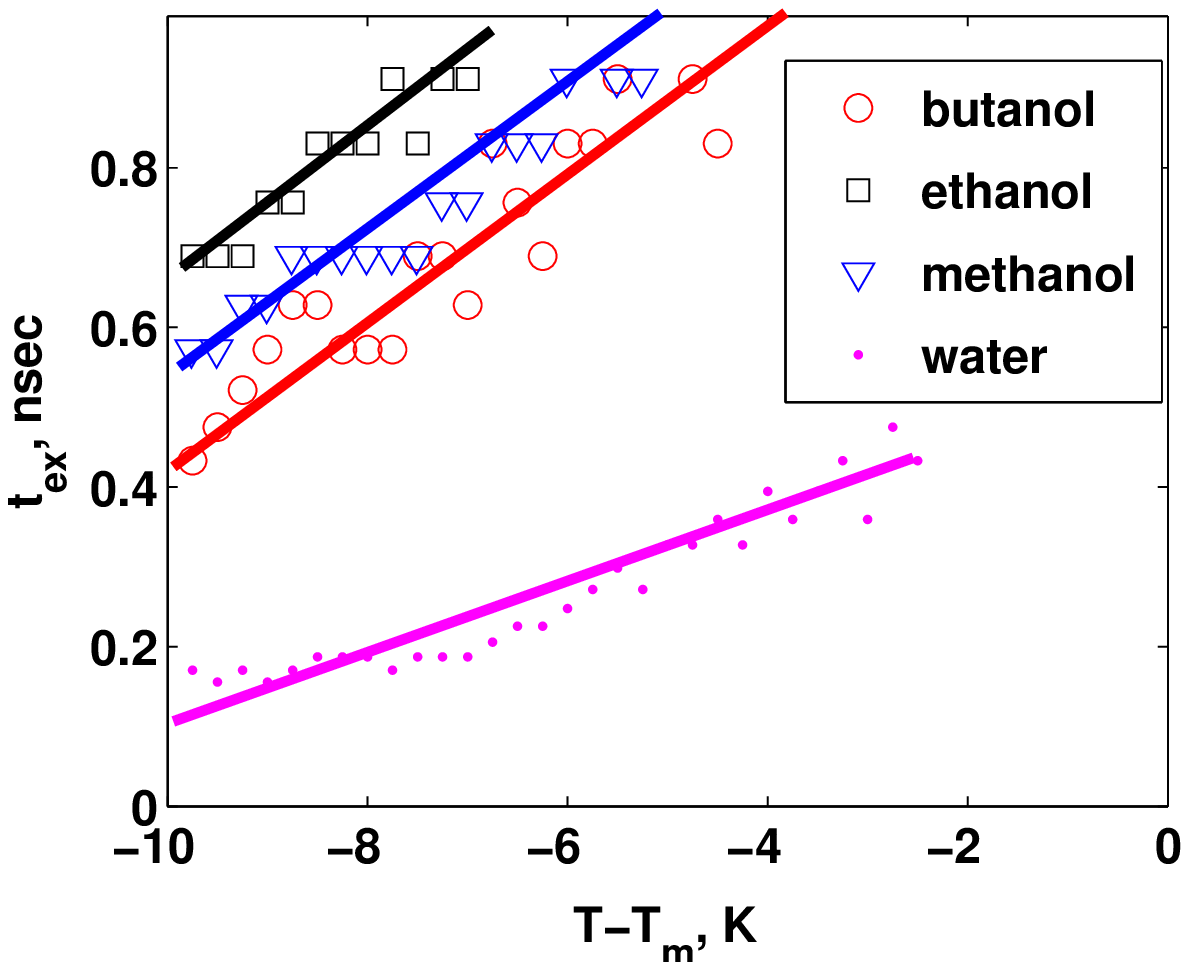} \includegraphics[scale=0.5]{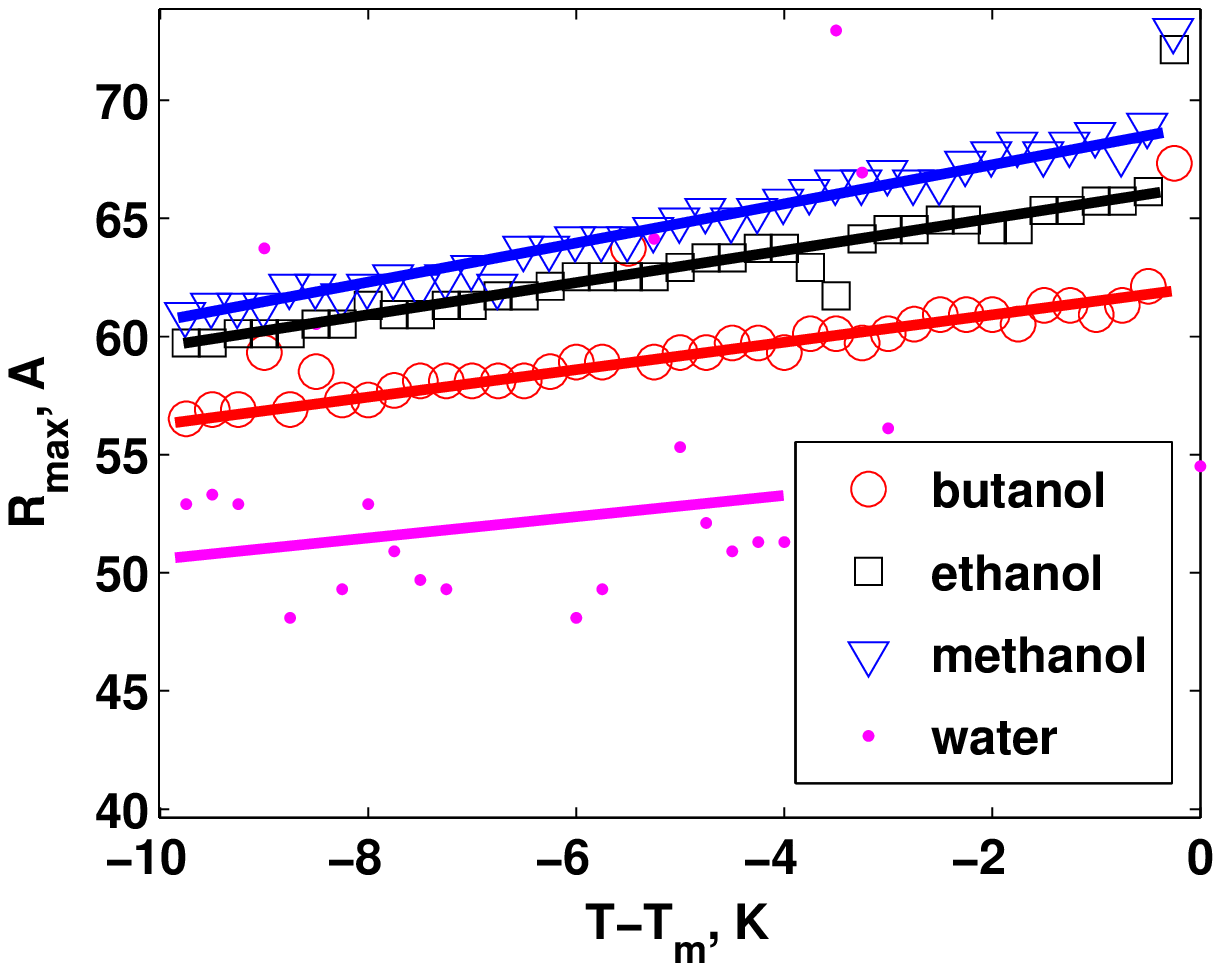}
\caption{Lifetime of the molten region $t_{m}$ (left) and its radius $R_m$ (right) for methanol, ethanol, butanol and water are displayed as a function of $T_b-T_m$.}
 \label{fig:features}
\vspace{-20pt}
\end{figure}

At the transition, the phase of the medium is deduced from the temperature of each spatial point, which can be higher than $T_{m}+\Delta T$ (liquid phase) or lower than $T_{m}-\Delta T$ (solid phase). Fig. \ref{fig:features} displays the maximum radius of the molten region, $R_{max}$, vs. $T_{\rm bulk}$ and the lifetime $t_{max}$ of the molten region (at $t>t_{max}$ temperature of an any point of the medium is below $T_m+\Delta T$).

\section{Positronium formation in condensed media}

\subsection{The Ore model}

The Ore model was proposed for interpretation of the Ps formation in gases \cite{Mog95}. It implies that the
``hot'' positron, e$^{+*}$, having excess kinetic energy, pulls out an electron from a molecule M, thereby
forming a Ps atom and leaving behind a positively charged radical--cation M$^{+\cdot}$:
\begin{eqnarray}
\rm e^{+*}+M &\to \rm Ps + M^{+\cdot}. \label{Eq3b}
\end{eqnarray}
This process is most effective when the e$^{+*}$ energy $W$ lies within the ``Ore gap'':
\begin{equation}  \label{Eq4}
 I_G-Ry/2 < W < W_{ex} ~~(\hbox{or } I_G).
\end{equation}
Here $I_G$ is the first ionization potential of the molecule, $W_{ex}$ is its electronic excitation threshold and $Ry/2=6.8$ eV is the Ps binding energy in a gas phase. If the positron energy is lower than $I_G-Ry/2$, e$^+$ cannot pick up an electron from a molecule. When $W>W_{ex}$, electronic excitations and ionizations dominate and Ps formation becomes less effective.

\begin{figure}[t]
\begin{center}
\epsfig{file=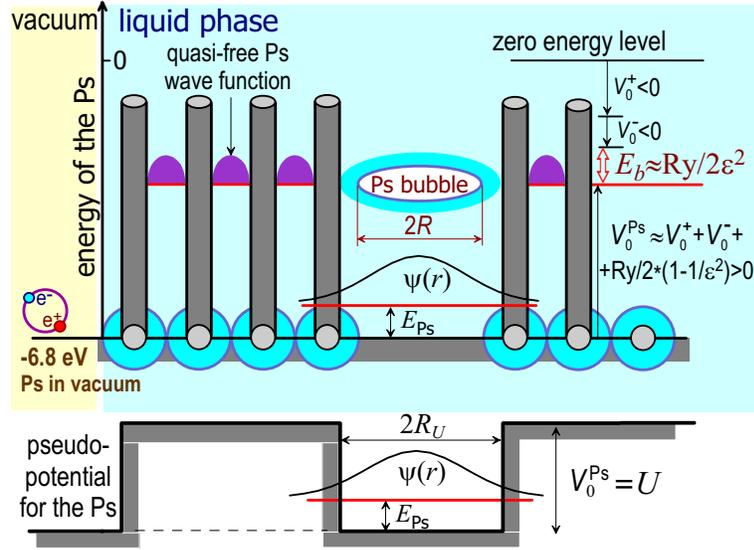, width=100mm}
\caption{Energy diagram of the Ps states. Energy of the Ps atom in vacuum is $-6.8$ eV. $V_0^{\rm Ps}$ is the Ps work function (the energy needed for Ps to enter the liquid without any molecular rearrangement). qf-Ps senses the repulsive potential from the cores of the molecules (for convenience, this potential is shifted down by $6.8$ eV). Being averaged, the latter may be represented as a step of height $V_0^{\rm Ps}=U$. $E_b\approx {\rm Ry}/2\varepsilon^2$ is a rough estimation of the qf-Ps energy in a dielectric continuum. Within some picoseconds, qf-Ps transforms into the bubble state with the free-volume radius, $R$, the center-of-mass wave function of the Ps, $\psi(r)$, and the ground-state energy $E_{\rm Ps}-Ry/2$ (for large bubbles). $R_U$ is the radius of the potential well, where Ps becomes localized.}
\label{Ps_states}
\end{center}
\end{figure}

\subsection{Quasifree Ps state}

In the condensed phase, because of the presence of molecules and lack of free space, the final Ps state differs from that in vacuum \cite{Ste02JCP}. We call this state quasifree positronium (qf-Ps). If we adopt a binding energy $E_b$ of qf-Ps in a dielectric continuum of roughly $\rm Ry/2\varepsilon^2$ (instead of just Ry/2 in vacuum), where $\varepsilon = n^2 \approx 2$ high-frequency dielectric permittivity is the square of the refractive index, it is seen that the Ore gap in a medium gets squeezed
\begin{equation} \label{Eq5}
 I_L - Ry/2\varepsilon^2 < W < W_{ex}^L
\end{equation}
and may even completely disappear because of a significant decrease in the binding energy. Here, $W_{ex}^L$ is the lower threshold of electronic excitations of a molecule in a liquid, $I_L=I_G - V_0^- -|U_p|$ is the liquid phase ionizing potential of a molecule ($V_0^-$ is the electron work function, or the energy of the ground state of electron in a medium and $U_p$ is the energy of polarization interaction of the positively charged ion with the medium).

In \cite{Cao98} it was suggested that in some polymers, because of the presence of large free-volume space, the final Ps state may not be quasifree: Ps would form immediately in one of the preexisting voids. If it did, the total energy gain would be larger and the Ore gap could exist. However, because the final Ps state has obviously zero translational momentum in contrast with the projectile ``hot'' positron, this reaction must be suppressed owing to the conservation law of momentum. e$^{+*}$ cannot promptly transfer its momentum to surrounding molecules because of the large difference in masses. Momentum relaxation as well as energy absorption during formation of the localized Ps in a cavity cannot be immediate and requires some time. Probably, we should expect the formation of qf-Ps in the bulk at first. Then this particle can lose energy and momentum before eventually ending up in a suitable cavity after a while.
\subsection{Recombination mechanism of Ps formation}

The above discussion points out the unique possibility for Ps formation in molecular media. This mechanism postulates that Ps is formed through the combination of the thermalized particles -- quasifree positron and one of the intratrack electrons:
\begin{equation}  \label{Eq6}
\rm e^+_{qf} + e^-_{blob} ~\to~ (e^+ \cdots e^-) ~\to ~\hbox{qf-Ps} ~\to ~\hbox{Ps in the bubble}.
\end{equation}
This reaction proceeds in the terminal part of the e$^+$ track (in the e$^+$ blob). If the positron is thermalized outside the blob, the only way for it to form Ps is to diffuse back and pick up one of the intrablob electrons. Otherwise it annihilates as a ``free'' positron. When e$^+$ picks up an electron, the ``initial'' separation between them must be comparable to the average distance $(4 \pi a_{bl}^3/3n_0)^{1/3} \approx 20$ \AA\ between intrablob species. The total energy of this pair is approximately the sum $V_0^+ + V_0^-$ of the e$^+$ and e$^-$ work functions. In comparison with the work functions, the binding energy of such a e$^+ \cdots$ e$^-$ pair is small (about 0.1 eV). However, the translational kinetic energies of the particles must be less than the binding energy, otherwise the pair will break up. So, just before Ps formation, the positron and track electrons must be almost thermalized.

\begin{figure}[t]
\centering
\includegraphics*[width=80mm]{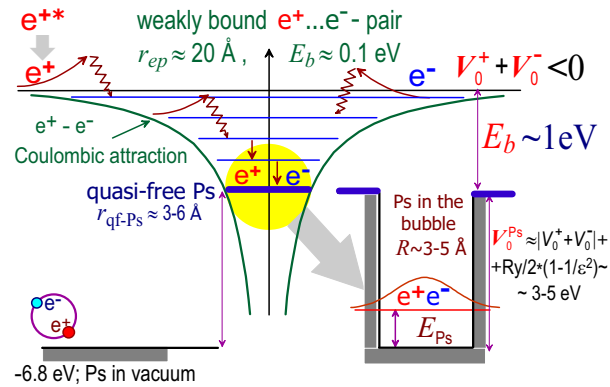}
\caption{Mechanism of the Ps formation. Here, for simplicity, the potential in which the Ps is localized, is shown as a rectangular well.} \label{F_Ps_form}
\end{figure}
However, such a weekly bound pair is not at the bottom of its energy spectrum, Fig. \ref{F_Ps_form}. The two particles approach each other (in average) and continue to release energy via excitation of molecular vibrations. Finally, the pair reaches an equilibrium state, which we term the quasifree Ps. Roughly, the binding energy of qf-Ps is $E_b \approx \rm Ry/2\varepsilon^2$, about 1 eV. In liquids, further energy gain of the pair is related with the rearrangement of molecules and appearance of some additional free space around Ps. e$^+$ and e$^-$ get closer, repelling molecules from their location and forming a Ps bubble state. A substantial decrease in their Coulombic energy is the driving force of this process.

Over the last 30 years the (re)combination mechanism has become extremely widespread \cite{Mog95,Ito88}.  It has been used to interpret numerous data on Ps chemistry, and explain variations of the Ps yields (from 0 to 0.7) in very different condensed media, where parameters of the Ore gap are practically the same. It provides a natural explanation to the changes in the Ps formation probability at phase transitions. Experimentally, the observed monotonic inhibition of Ps yields (practically down to zero) in solutions of electron acceptors contradicts the Ore model, but inserts well in the recombination mechanism. It explains the anti-inhibition effect, including experiments on Ps formation in moderate electric fields in pure liquids and mixtures.

There are two models which utilize this mechanism, the spur model \cite{Mog74} and the blob model (diffusion-recombination model) \cite{Bya74,Bya76}. In spite of the fact that both models answer the question about the Ps precursor in the same way, they differ as to what constitutes the terminal part of the e$^+$ track and how to calculate the probability of the Ps formation \cite{Ste02JCP}.

\section{Intratrack reactions in the positron blob}

\begin{figure}[!t]
\begin{minipage}[h]{70mm}
\includegraphics[width=70mm]{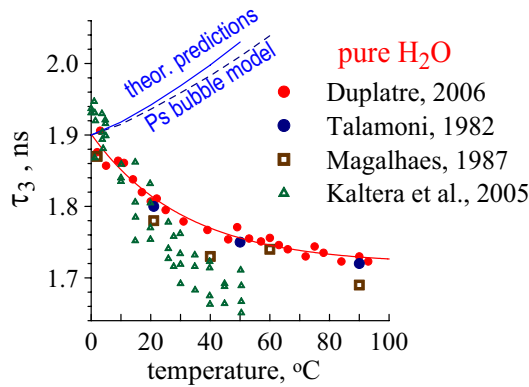}
\end{minipage}
\begin{minipage}[h]{80mm}
\caption{Temperature dependence of the lifetime, $\tau_3$, of the long-lived component in pure water: experimental data derived from a 3-exponential analysis and corresponding theoretical predictions: finite potential well (solid line), infinite potential well (dashed line) \cite{09Ste_PSS}.} \label{F_tau3}
\end{minipage}
\end{figure}

The interaction of the Ps atom with primary radiolytic products, formed in the terminal part of the positron track (in the e$^+$ blob) due to e$^+$ ionization slowing down, is feature inherent to the positron spectroscopy of any molecular medium and must be taken into account in interpreting the annihilation spectra. Many intratrack products are strong oxidizers and/or radicals (like $\rm H_3O^+$ radical-cations and OH-radicals in water) and their initial concentration in the e$^+$ blob is not small (up to 0.05 M). The average distance between intrablob species is comparable with the diffusion displacement of the Ps bubble before e$^+$ annihilation, so the contribution of diffusion-controlled oxidation and ortho-para conversion reactions is quite possible.

Neglecting these reactions leads to obvious contradictions. Fig. \ref{F_tau3} shows the $T$-dependence of the lifetime ($\tau_3$) of the long-lived component of PAL spectra in pure water (roughly, $\tau_3$ is the pick-off lifetime of ortho-positronium). Experimental data from different authors reasonably agree, but all are in strong contradiction with the theoretical expectations based on any Ps bubble model \cite{09Ste_MSF}: with increasing $T$, the surface tension coefficient decreases, so the size of the Ps bubble should increase, which should lead to an increase in $\tau_3$ with $T$, in sharp contradiction to the PAL data.

One most informative way to investigate Ps intratrack reactions is to study the temperature variation of the PAL spectra. We have chosen liquid water to initiate this investigation, because radiolytic processes in water have been studied for decades and are well known. In addition, the Ps bubble growth proceeds very fast in water ($\lesssim 10$ ps) and therefore may be considered as an instantaneous process. An opposite situation is true in glycerol at $T\lesssim 50$ $^\circ$C, where Ps interaction with intratrack radiolytic products and formation of the Ps bubble state may occur simultaneously \cite{09Ste_MSF_Ps_bub}. Because all these processes strongly depend on temperature, it is worth to consider two extreme cases, the cases of ``high'' and ``low'' temperatures.

\subsection{High temperature region}

This region may be defined through the following conditions:

1) the Ps bubble formation time, $t_{bubble}(T) \approx R\eta/\sigma$ \cite{Mik05} is short as compared to the time resolution of a PAL spectrometer, $\sim 0.1$-0.2 ns. Here, $R\approx 3$-6 \AA\ is the radius of the Ps bubble, $\eta$ and $\sigma$ are viscosity and surface tension coefficients of a liquid medium. Experimental observation of the bubble growth is impossible in this case because its equilibrium value is reached too fast (less than 0.1 ns).

\begin{figure}[t]
\begin{minipage}[h]{55mm}
 \epsfig{file=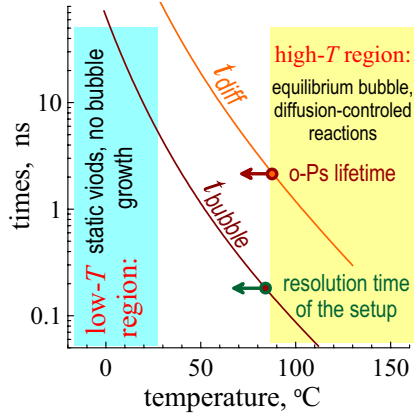, width=55mm}
\end{minipage}
\begin{minipage}[h]{90mm}
\caption{Temperature dependences of the Ps bubble formation time $t_{bubble}$ and typical time $t_{diff}$
of the diffusion-controlled intrablob reactions in glycerol.}  \label{F2}
\end{minipage}
\end{figure}

2) the diffusion length $\sqrt{6(D_{\rm Ps}+D_i)\tau_3}$ of Ps and radiolysis products (subscript $i$) in the e$^+$ blob during the ortho-Ps lifetime, $\tau_3$ (about few nanoseconds), must exceed the average distance between intrablob particles $\bar r \approx 10$-20 \AA. This can be expressed through the condition $t_{diff}<\tau_3$, where $t_{diff} = \bar r^2 / 6(D_i+D_{\rm Ps})$; on the basis of the Stokes-Einstein relationship, $D_i\approx D_{\rm Ps}\approx k_B T/6\pi\eta(T) R_i$, $R_i$ is a few \AA. In glycerol both conditions are satisfied at the same temperature region $T>90$ $^\circ$C (Fig. \ref{F2}), but in liquid water they are fulfilled at any $T$ from 0 to 100 $^\circ$C .

In the high-$T$ region the influence of intrablob reactions is important. The radiolytic processes initiated through ionizations induced by the slowing down of fast e$^{+*}$ may be represented through the following basic reactions:

ionization:
\begin{equation} \label{R1}
\rm e^{+*} + RH \to e^- + \dot RH^+ + e^+,
\end{equation}

ion-molecule reaction:
\begin{equation} \label{R1a}
\rm \dot RH^+ + RH \to RH_2^+ + \dot R,
\end{equation}

ion-electron recombination:
\begin{equation} \label{R2}
\rm e^- + \dot RH^+ \to  \cdot \dot RH^*_{\rm tripl}, ~ RH^*_{\rm singl};
\end{equation}
\begin{equation} \label{R4}
\rm e^- + RH_2^+  \to RH + \dot H \to \dot R + H_2,
\end{equation}

electronic deexcitation:
\begin{equation} \label{R3}
\rm \cdot \dot RH^*_{\rm tripl} \stackrel{+ \rm RH}{~\longrightarrow~} 2\dot R+H_2,
 \qquad
RH^*_{\rm singl} \to RH.
\end{equation}
From reactions (\ref{R1}-\ref{R4}) one can infer that the total number of $\rm \dot R$ in the e$^+$ blob varies weakly in time and is approximately equal to the initial number, $n_0$, of ion-electron pairs in the blob. However, their spatial distribution broadens in time due to out-diffusion. The main effect of ions and radicals on Ps is oxidation and spin conversion reactions. If we denote for simplicity all these species ($\rm \dot RH^+$, $\rm RH_2^+$, $\rm \dot R$, $\rm \dot H$) by the same symbol, $\rm \dot R$, and term them ``radicals'', their reactions with Ps may be written as:

Ps oxidation:
\begin{equation} \label{R5}
\rm Ps + \dot R \to e^+ + R^-,
\end{equation}

ortho-para conversion:
\begin{equation} \label{R6}
\hbox{para- or ortho-Ps } + \rm \dot R \to
 \frac{1}{4} \hbox{para-Ps} + \frac{3}{4} \hbox{ortho-Ps} + \dot R.
\end{equation}
In the high temperature region these Ps reactions are mostly diffusion-controlled. With the help of the nonhomogeneous chemical kinetics approach (the blob model, \cite{09Ste_MSF,03Ste_PRCPP}) reactions (\ref{R1}-\ref{R6}) may be described in terms of the following equations on the concentrations of different intratrack species:
\begin{equation} \label{Eq9}
\frac{\partial c_i(r,t)}{\partial t} = D_i \Delta c_i - k_{ij} c_i c_j - \lambda_i c_i,
 \qquad
c_i(r,0)=c_i^0 \frac{\exp(-r^2/a_i^2)}{\pi^{3/2} a_i^3}.
\end{equation}
where $c_i(r,t)$ is the concentration of the intrablob species of the $i$-th type (including all the positron states: e$^+$, para-Ps, ortho-Ps), $k_{ij}$ is the rate constant of the reaction between the $i$ and $j$ reactants and $\lambda_i$ is the decay rate of the $i$-th particles including possible annihilation.

According to the Smoluchowski approach, the reaction rate constant $k_{\rm Ps,R}$ (let us take as an example reactions (\ref{R5}-\ref{R6})) may be written as follows:
\begin{equation}\label{Eq7}
k_{\rm Ps,R}(T,t) = 4\pi D_{\rm Ps,R}(T) R_{\rm Ps,R} \cdot
 \left[ 1 + \frac{R_{\rm Ps,R}}{\sqrt{\pi D_{\rm Ps,R}t}} \right],
\end{equation}
where $R_{\rm Ps,R} = R_{\rm Ps}+R$ is the reaction radius of the $i + \rm Ps$ chemical reaction, $R_i$ is approximately the geometric radius of $i$-th reactant, $R_{\rm Ps} \approx R_U +1/\varkappa$ is the radius of delocalization of the Ps wave function ($1/\varkappa$ is the under-barrier penetration length of the Ps center-of-mass wave function; here, we consider Ps as a point particle in a potential well).

The temperature dependence of the rate constant arises through the diffusion coefficients of the reagents, $D_{\rm Ps,R}(T) = D_{\rm Ps}+D_{\rm R}$. According to the Einstein relationship and the Stokes formula, the diffusion coefficients may be expressed as:
\begin{equation}\label{Eq7a}
D_{\rm R} = \frac{T}{6 \pi \eta(T) R}, \qquad D_{\rm Ps} = \frac{T}{4 \pi \eta(T) R_{\rm Ps}},
\end{equation}
where $\eta$ is the viscosity of the medium and $R$ and $R_{\rm Ps}$ are the hydrodynamic radii of $\rm \dot R$ particle and Ps bubble. Although the applicability of these expressions on the atomic scale might be questionable, our studies in neat water \cite{09Ste_MSF,09Ste_PSS} have shown that it remains valid within a reasonable accuracy. Usually the viscosity $\eta(T)$ drops down approximately exponentially with $T$.

The knowledge of the surface tension of the liquid together with the Ps bubble model allow one to estimate the equilibrium radius of the Ps bubble and calculate the pick-off annihilation rate $\lambda_{po}(t,T)$, which is of utmost importance to interpret the PAL spectra:
\begin{equation}\label{Eq8}
\lambda_{po}(t,T) = \lambda_p \cdot P(R_U(t,T)).
\end{equation}
Here, $\lambda_p$ is the ``free'' positron annihilation rate and $P(R_U)$ is the underbarrier penetration probability of e$^+$ into a bulk of the liquid (into space containing molecular electrons). Its calculation is discussed below.

\subsection{Low temperature region}

In this region, the Ps bubble formation time is larger than the e$^+$ and Ps annihilation lifetimes. On a scale of 1 ns, the size of a preexisting void in which a quasi-free Ps has been localized does not change, since $t_{bubble} \gg 1$ ns. This situation is similar to that existing in polymers.  Typically at these $T$ the diffusion time $t_{diff}$ is much larger than o-Ps lifetime ($\sim 1$ ns), so one may completely neglect the diffusion motion of radiolytic products. For example in glycerol it happens at $T < 30$ $^\circ$C, Fig.
\ref{F2}.

To describe correctly the PAL spectra, one should average the pick-off annihilation kinetics over the size distribution of the preexisting voids in which Ps localization takes place. This can be done by using the theory of free volume entropy fluctuations \cite{Bur}, which claims that every molecule possesses some free volume $v_i$ with the probability
\begin{equation} \label{Eq10}
\propto \exp(-v_i/v_F), \qquad v_F(T) \approx v_{\rm WS}(T) - v_{\rm vdW},
\end{equation}
where $v_F$ is a difference between the volume of the Wigner-Seitz cell and van-der-Waals volume of a molecule. Because of the approximate character of this expression, $v_F$ may be considered as an adjustable parameter derivable by fitting the spectra in the low temperature region.

If $v$ is the volume of a preexisting void in which the Ps atom can localize, the probability to find such a void is proportional to $\exp(-v/v_F)$. Neglecting the influence of any intratrack reactions (because of low-$T$ and not discussing their possible tunneling nature), the calculation of the Ps pick-off annihilation kinetics reduces to averaging the exponents that relate to the pick-off annihilation of individual Ps atoms located at different voids:
\begin{equation} \label{Eq11}
 \left\langle e^{-\lambda_{po}(R_U)t} \right\rangle =
 \int_{v_{min}}^\infty e^{-\lambda_{po}(R_U)t} \cdot e^{-v/v_F} \cdot \frac{dv}{v_F},
\qquad v = \frac{4\pi R_U^3}{3}.
\end{equation}
Here, $v_{min}$ is the volume of the minimal cavity which may trap Ps (section \ref{non-point}).

\subsection{Intermediate temperatures}

The most complex task in the PAL data processing refers to the intermediate temperature region, where the bubble formation time is comparable with the e$^+$ lifetime in the medium. qf-Ps localizes in one of the preexisting voids, which further slightly increases in size; the time at which the equilibrium volume is reached depends on $T$. Obviously, the quantum-mechanical pressure exerted by Ps on the wall and thus, the growth of the void, cease when e$^+$ annihilates.

At these temperatures, the Ps fate may also depend on the intratrack reactions. Therefore, the corresponding chemical kinetic equations must be solved by assuming that Ps initially localizes in a preexisting void of a given radius, which will increase with time until annihilation; the process must be averaged over the size distribution of the preexisting voids \cite{09Ste_MSF_Ps_bub}.

\subsection{Interpretation the PAL spectra in liquid water at different $T$}

To solve the problem with a puzzling temperature behavior of the long-lived component in water, Fig. \ref{F_tau3}, we proposed to reject the conventional exponential deconvolution of the spectra and explicitly take into account the intratrack processes mentioned above \cite{09Ste_MSF,11Ste_MSF}. The decrease of $\tau_3$ vs. $T$ was ascribed to the increasing efficiency of the oxidation (\ref{R5}) and ortho-para-conversion (\ref{R6}) reactions between Ps and intratrack radicals (mainly OH-radicals).

All parameters included in the model have a clear physical meaning: the Ps oxidation reaction radius $R_{ox}\approx R_{\rm Ps}+R_{\rm OH}$ which enters the rate constant of the Ps oxidation reaction (\ref{R5}), relative contact density $\eta_c$ in the Ps bubble state, free positron annihilation rate, the Ps formation rate constant and some others (5 adjustable parameters in total).

\begin{figure}[t]
 \centering
 \includegraphics[width=110mm]{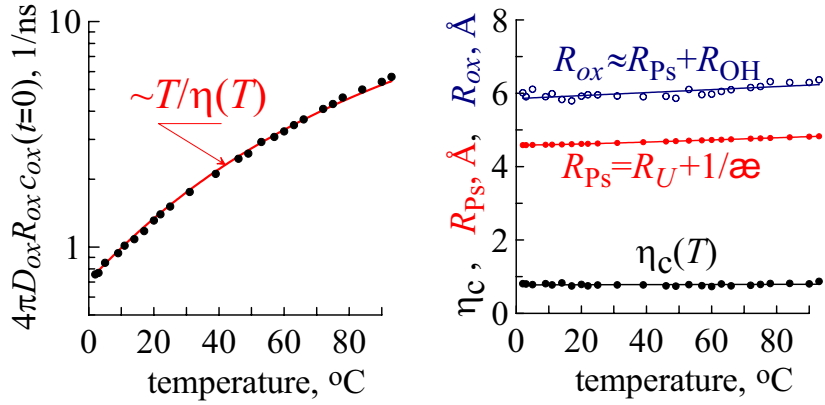}
\caption{{\bf Left:} Temperature dependence of the Ps oxidation reaction rate constant (at an infinite time, $t \to \infty$) multiplied by the ``initial'' concentration of oxidizers $c_{ox}(0)$ in the e$^+$ blob (OH and $\rm H_3O^+$, see \cite{09Ste_MSF,11Ste_MSF} for details). $D_{ox}$ is the sum of $D_{\rm Ps}$ and the diffusion coefficient of the oxidizer (OH radical, $D_{\rm OH}\approx 1.5 \cdot 10^{-5}$ cm$^2$/s). The obtained dependence agrees well with the Stokes-Einstein law, $\propto T/\eta(T)$.
{\bf Right:} $T$-dependences of the oxidation reaction radius, $R_{ox,\rm Ps}=R_{ox}+R_{\rm Ps}$, the Ps radius, $R_{\rm Ps}=R_U +1/\varkappa$ and the relative contact density parameter $\eta_c$ in Ps. $R_{ox}$ is the radius of the oxidizer (approximately, the geometric radius of the OH radical, 1.4 \AA).}  \label{F_k_ox}
\end{figure}

It is seen that the blob model explains well the experimental data in a wide range of temperatures and magnetic fields. It does not lead to contradictions with known radiation chemistry data. The agreement between theory and experiment became even better when taking into account the time dependence of the Ps reaction rate coefficients. The temperature dependence of the Ps diffusion controlled reaction rate constants agrees with the Stokes-Einstein law. As we have obtained. Good fitting of the PAL data is obtained when the Ps hydrodynamic radius is equal to the radius $R_{\rm Ps}=R_U +1/\varkappa$ of the Ps bubble (here $1/\varkappa \approx 0.5$ \AA\ is the underbarrier penetration depth of the Ps wave function).

\section{Ps bubble models}  \label{non-point}

Typical lifetimes of a para-positronium atom in condensed medium are about 130-180 ps. They are close to the p-Ps lifetime in vacuum (125 ps). The ortho-positronium lifetime in a medium is considerably shorter (about 100 times; some ns) in comparison with that in vacuum. This is due to the so-called pick-off process -- prompt 2$\gamma$-annihilation of the e$^+$, composing Ps atom, with one of the nearest e$^-$ of surrounding molecules, whose spin is antiparallel to the e$^+$ spin. Just this property turns Ps into a nanoscale structural probe of matter. The theoretical task consists in calculating the pick-off annihilation rate $\lambda_{po}$, i.e. in relating $\lambda_{po}$ with such properties of the medium like surface tension, viscosity, external pressure and size of the Ps trap.

Originally, to explain the unexpectedly long lifetime of the ortho-Ps atom in liquid helium R.Ferrel \cite{Fer57} suggested that the Ps atom forms a nanobubble around itself. This is caused by a strong exchange repulsion between the o-Ps electron and electrons of the surrounding He atoms. Ferrel approximated this repulsion by a spherically symmetric potential barrier of radius $R_\infty$. To estimate the equilibrium radius of the Ps bubble he minimized the sum of the Ps energy in a spherically symmetric potential well, i.e. $\pi^2\hbar^2/4mR_\infty^2=\frac{\rm Ry}{2} (\pi a_B/R_\infty)^2$, Ry=13.6 eV, and the surface energy, $4\pi R_\infty^2 \sigma$, where $\sigma$ is the macroscopic surface tension coefficient. The following relationship is hereby obtained for the equilibrium radius of the bubble:
\begin{equation}\label{9_2}
\frac{\pi^2 a_B^2}{R_\infty^2} {\rm Ry} + 4\pi R_\infty^2 \sigma ~\leftrightarrow  ~ \hbox{min over} ~R_\infty
 ~~~\Rightarrow ~~~
R_\infty = a_B \left( \frac{\pi {\rm Ry}}{8\sigma a_B^2} \right)^{1/4}.
\end{equation}

\subsection{The Tao-Eldrup model}

\begin{figure}[t]
\centering
\begin{minipage}[h]{35mm}
 \epsfig{file=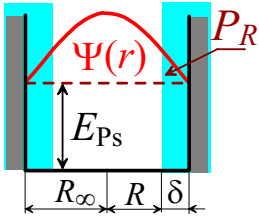, width=35mm}
\end{minipage}
\begin{minipage}[h]{100mm}
\caption{Tao-Eldrup model of Ps atom in a liquid phase. It is assumed that Ps is confined in an infinite spherically symmetric potential well of radius $R_{\infty}$. $\Psi(r)$ is the center-of-mass wave function of Ps. $R$ is the free volume radius of the Ps bubble. $\delta$ is the penetration depth of the molecular electrons into the Ps bubble.} \label{F_TF}
\end{minipage}
\end{figure}

Ferrel's idea got further development in the studies of Tao \cite{Tao72} and Eldrup et al. \cite{Eld81}. They considered the Ps atom as a point particle in a liquid, i.e. in a structureless continuum, Fig. \ref{F_TF}. The repulsive Ps-liquid interaction was approximated by a rectangular infinitely deep spherically symmetric potential well of radius $R_{\infty}$. In such a well, the wave function of a point particle has the following standard
expression:
\begin{equation}\label{9_3}
\Psi (0\le r \le R_\infty) = \frac{\sin(\pi r/R_\infty)}{\sqrt{2\pi R_\infty} ~ r},
 \qquad
\Psi (r\ge R_\infty) = 0.
\end{equation}
 \noindent
Here, $r$ is the Ps center-of-mass coordinate. Because the Ps wave function equals to zero at the bubble radius (and outside), there is no e$^+$ overlapping with outer electrons of a medium. So, pick-off annihilation is absent. To overcome this difficulty it was postulated that molecular electrons, which form a ``wall'' of the Ps bubble, may penetrate inside the potential well. This results in the appearance of a surface layer of thickness $\delta = R_{\infty}-R$ having the same average electron density as in the bulk. As a result, the pick-off annihilation rate $\lambda_{po}$ becomes non-zero. It is proportional to the e$^+$ overlapping integral with the electrons inside the bubble:
\begin{equation}\label{9_5}
\lambda_{po} = \lambda_+ P_R, \qquad P_R = \int_{R}^{R_{\infty}} |\Psi(r)|^2 4\pi r^2 dr
=\frac{\delta}{R_\infty} - \frac{\sin (2\pi \delta / R_\infty)}{2\pi }.
\end{equation}
This is the well-known Tao-Eldrup formula. Here, $\lambda_+\approx 2$ ns$^{-1}$ is the e$^+$ annihilation rate in an unperturbed medium (it is proportional to Dirac's 2$\gamma$-annihilation cross-section and the number density of valence electrons). The thickness $\delta$ of the electron layer is an empirical parameter, which may have different values in various media.

\begin{figure}[t!]
\begin{minipage}[h]{75mm}
  \includegraphics[width=75mm]{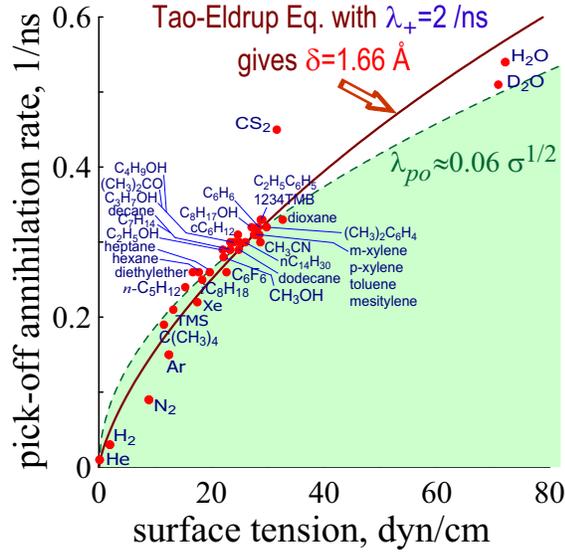}
\end{minipage}
\begin{minipage}[h]{75mm}
\caption{Dependence of the experimental pick-off annihilation rates \cite{Ste02JSC} vs. surface tension in different liquids.
The solid curve shows the correlation given by the Tao-Eldrup at
$\lambda_+ = 2$ ns$^{-1}$ and optimal value $\delta = 1.66$ \AA\ (obtained from fitting of these
data by means of Eq. (\ref{9_5})). The dashed curve illustrates the simplest approximation
$\lambda_{po} \propto \sigma^{1/2}$.} \label{Lpo_sig}
\end{minipage}
\end{figure}

Substituting Eq. (\ref{9_2}) for $R_\infty$ into Eq. (\ref{9_5}), one obtains the relationship between $\lambda_{po}$ and $\sigma$ with one adjustable parameter, $\delta$. It may be easily obtained by fitting experimental pick-off annihilation rates with the relationship (\ref{9_5}), Fig. \ref{Lpo_sig}. Thus we obtain $\delta \approx 1.66$ \AA. Eq. (\ref{9_5}) with this value of $\delta$ is widely used for recalculation of the observed pick-off annihilation rate into the free volume $4\pi R^3/3$ of the cavity, where Ps atom resides and annihilates.

\subsection{Further development of the Ps bubble models. ``Non-point'' positronium}

Along with the development of the ``infinite potential well'' Ps bubble model, another approach based on the finite potential well approximation was also elaborated \cite{Stew59,Bych61,Roel67,Dau00,Ste02JSC}. However in both approaches, the Ps atom was approximated by a point particle. This leads to a significant simplification, but it is not justified from a physical viewpoint, because:

1) the size of the localized state of Ps (size of the Ps bubble) does not significantly exceed the distance between e$^+$ and e$^-$ in Ps;

2) during the formation of the Ps bubble there is a substantial variation of the Ps internal energy (particularly of the Coulombic attraction of e$^+$ and e$^-$), which is completely ignored in the ``point-like'' Ps models. In a vacuum or in a large bubble, the internal energy of Ps tends to $-{\rm Ry}/2=-6.8$ eV. In a continuous liquid (no bubble) with the high-frequency dielectric permittivity $\varepsilon \approx n^2$ ($n\approx 2$-3 is the refractive index) the energy of the Coulombic attraction between e$^+$ and e$^-$ decreases in absolute value by a factor $\varepsilon^2 \approx 4$-9. The same takes place with the total Ps binding energy, which tends to the value $-{\rm Ry}/2\varepsilon^2 \approx -(1$-1.7) eV (this is a simple consequence of the scaling $e^2 \to e^2/\varepsilon$ of the Schr\"odinger equation for Ps atom). Thus, the change in the Ps internal energy during Ps formation may reach 5 eV. Obviously, this represents an important contribution to the energetics of Ps formation. The aim of the present work is towards a more accurate estimation of this contribution, that has not been done yet.

There is only a small number of papers where the consequences of the finite size of Ps are discussed in application to positron annihilation spectroscopy. To calculate $\lambda_{po}$, the Kolkata group \cite{Dut02} suggested to smear the Ps atom over the relative e$^+$-e$^-$ coordinate exactly in the same way as it is in a vacuum. Such an approach is valid for rather large bubbles. However, they do not discuss the variation of the internal Ps energy.

In \cite{Seeg03} the Ps atom is considered as a finite sized e$^-$e$^+$ pair, but the variation of the Coulombic interaction because of dielectric screening is not discussed. It was assumed that e$^-$ is confined in an infinite potential well and e$^+$ is bound to it by means of the Coulombic attraction. The wave function of the pair was taken as a series of orthogonal polynomials, their weights being determined from a minimization procedure of the total energy of the pair.

\subsection{Hamiltonian of e$^+$e$^-$ pair in a medium}

Let the e$^+$e$^-$ pair (Ps atom) have already formed in a liquid a nanobubble (spherical cavity; Ps bubble) of radius $R$ (the onset of coordinates is taken at the center of the bubble, Fig. \ref{Bub_coord}). Together with the molecules surrounding the e$^+$e$^-$ pair, one has to deal with a quite intricate many-body problem with a complex hamiltonian. We reduce it to the following form:
\begin{equation}\label{2_1}
H \approx -\frac{\hbar^2 (\Delta_+ + \Delta_-)}{2m_e} + U({\bf r}_+) + U({\bf r}_-)
 - U_c({\bf r}_+, {\bf r}_-, R, \varepsilon).
\end{equation}
\noindent
 \begin{floatingfigure}[h!]{35mm} 
  \noindent \includegraphics[width=30mm]{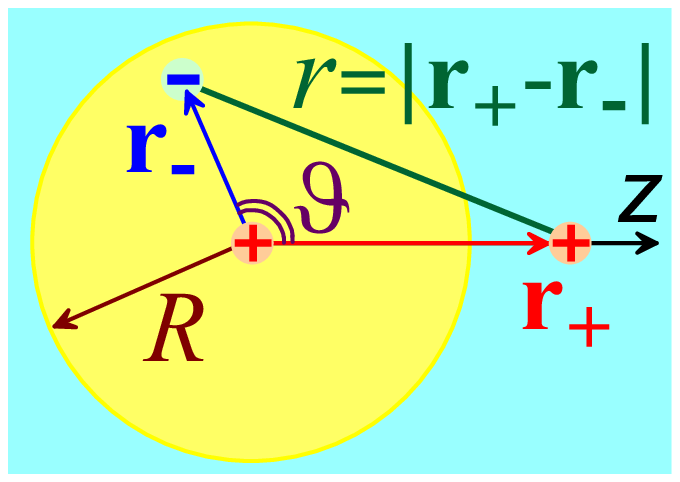}
  \label{Bub_coord}
 \end{floatingfigure}
\noindent Terms with Laplacians $\Delta_+$ and $\Delta_-$ over ${\bf r}_+$ and ${\bf r}_-$ (e$^+$ and e$^-$ coordinates) stand for the kinetic energies of the particles. $U({\bf r}_+)$ and $U({\bf r}_-)$ describe the individual interaction of e$^+$ and e$^-$ with the medium. For them we adopt the following approximation:
\begin{equation}\label{2_2}
U(r_+) = \left\{ \begin{tabular}{ll}
						  0 , & $r_+<R$,\\
						 $V_0^+$, & $r_+>R$,\\
						 \end{tabular}    \right.   \qquad
U(r_-) = \left\{ \begin{tabular}{ll}
						  0 , & $r_-<R$,\\
						 $V_0^-$, & $r_->R$.\\
						 \end{tabular}    \right.
\end{equation}

\noindent Here, $V_0^+$ and $V_0^-$ are the e$^+$ and e$^-$ work functions, respectively ($V_0$ is a commoner notation for the electron work function). The work function is usually introduced as the energy needed for an excess particle to enter the liquid without any rearrangement of its molecules and to stay there in a delocalized state, having no preferential location in a bulk. One may say that $V_0^+$ and $V_0^-$ are the ground state energies of the quasifree e$^+$ and e$^-$, because their energies at rest after having been removed from the liquid to infinity are defined to be zero.

\begin{table}
\caption{Electron work function for different liquids at room temperature \cite{HRC91}}  \label{V0}
\bigskip
 \centering
\begin{tabular}{|lc||lc|}
\hline
 Liquid             & $V_0^-$, eV  & Liquid        & $V_0^-$, eV \\
\hline
 helium; 4.2 K       &         1.3  & benzene        &-0.14 \\
 n-dodecane          &         0.2  & isooctane      &-0.17 \\
 n-decane            &         0.18 & toluene        &-0.22 \\
 n-heptane           &         0.12 & neopentane     &-0.38  \\
 n-hexane            &         0.1  & MeOH, EtOH, PrOH &-0.4\\
 nitrogen; 77.3 K    &         0.05 & xenon; 170 K   &-0.57  \\
 n-pentane, c-hexane &         0.01 & water          &-1.2   \\
 argon; 86.4 K       &         0    &                &       \\
\hline
\end{tabular}
\end{table}

$V_0^-$ consists of 1) the e$^-$ kinetic energy, arising from its exchange repulsion from the ``core'' electrons of molecules (atoms), and 2) the energy due to the polarization interaction of e$^-$ with the medium.\footnote{In case of e$^+$ the kinetic contribution to $V_0^+$ is due to the Coulombic repulsion from the nuclei (the exchange repulsion is absent).} According to the theory of the quasifree electron \cite{Spr68}, this polarization interaction may be estimated as a sum of two parts: a) interaction of the e$^-$ with the molecule where it resides, $U_-^{int}$ (to calculate $U_-^{int}$ the electron is considered as an electron cloud smeared over the molecule), and b) interaction of the e$^-$ with all the other molecules, $U_-^{out}=(1-1/\varepsilon)e^2/2R_{\rm WS}$, (this expression is similar to the well-known Born formula for the electron solvation energy).

Experimental values for $V_0^-$ are known for many liquids (Table \ref{V0}). Because of a lack of experimental data on the e$^+$ work functions, we shall admit that they are approximately the same as for e$^-$: $V_0^+ \approx V_0^-$. So we may conclude that $|V_0^+ +V_0^-|\lesssim 1$ eV. Thus $|V_0^+ +V_0^-|$ is less than the variation of the internal energy of the pair, $\approx {\rm Ry}(1-1/\varepsilon^2)/2 \approx 5$ eV, related with the variation in the dielectric screening of the e$^+$-e$^-$ attraction in the bubble formation process.

Note that the use of Eqs. (\ref{2_2}) for the potential energies of the e$^+$ and e$^-$ interaction with the medium, assumes that the polarization interaction remains the same whether e$^+$ and e$^-$ (both in the quasi-free states) are well separated or form the quasi-free Ps atom. Since, for distances larger than the size of a molecule, qf-Ps is nearly an electrically neutral particle, the contributions $U_-^{out} \approx U_+^{out}$, which come from a long-range polarization interaction of the quasi-free e$^+$ and e$^-$ with the medium, should be absent in $U(r_-)+U(r_+)$ in Eq. (\ref{2_1}). Therefore, it is reasonable to consider at least two cases: 1) when the above mentioned polarization correction is neglected and $V_0^- + V_+ \to 0$ and 2) when the terms $U_-^{out}\approx U_+^{out} \approx -1$ eV are subtracted from the work functions and $V_0^- + V_0^+ \to 2$ eV. Both cases are considered below.

\begin{figure}
\centering
 \includegraphics[width=120mm]{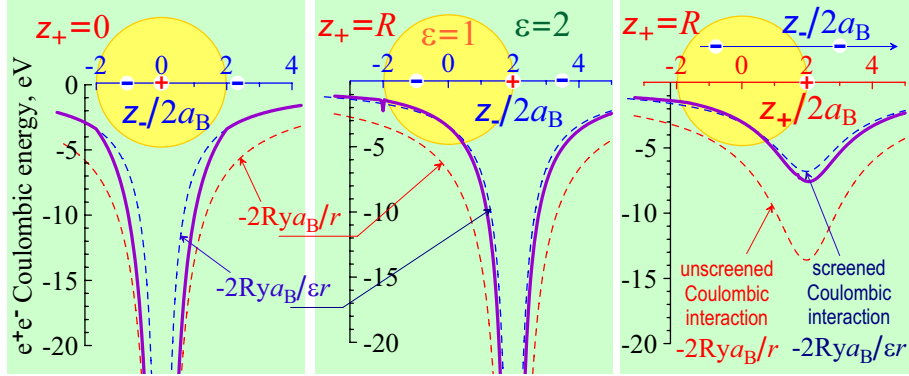}
\caption{Dependence of the e$^+$-e$^-$ Coulombic interaction energy for different locations of e$^+$ and e$^-$ around the bubble (here we adopt the radius of the bubble $R$ equal to $4a_B \approx 2$ \AA). $z_+$ and $z_-$ are the e$^+$ and e$^-$ displacements from the center of the bubble along the $z$-axis. The dashed curves describe the unscreened (red) and completely screened ($\varepsilon$ times less; blue) Coulombic energies between e$^+$ and e$^-$.}\label{BaT_Uc}
\end{figure}

In Eq. (\ref{2_1}) $U_c$ stands for the Coulombic interaction between e$^+$ and  e$^-$ in a polarizable medium. Assuming that the medium has the dielectric permittivity $\varepsilon$ of the bulk and a spherical cavity of radius $R$ (inside the cavity $\varepsilon=1$), one may calculate $U_c$ by solving the Poisson equation. Denoting the e$^+$ and e$^-$ coordinates as ${\bf r}_+$ and ${\bf r}_-$, $U_c$ may be written in the form of the following series via the Legendre polynomials $P_l(x=\cos \theta)$ \cite{Bat02}:
$$
\frac{U_c(r_+<R,r_-<R)}{\rm Ry} = \frac{2a_B}{r} -\left( 1-\frac{1}{\varepsilon} \right)\frac{2a_B}{R}
 \left(1+\sum_{l=1}^\infty \frac{(1+l)P_l(x)}{1+l+l/\varepsilon} \cdot \frac{r_+^l r_-^l}{R^{2l}}\right);
$$
$$
\frac{U_c(r_+<R,r_->R)}{\rm Ry} = \frac{2a_B}{\varepsilon r_-}
	 \left( 1+\sum_{l=1}^\infty \frac{(1+2l)P_l(x)}{1+l+l/\varepsilon} \cdot \frac{r_+^l}{r_-^l} \right);
$$
$$
\frac{U_c(r_+>R,r_-<R)}{\rm Ry} = \frac{2a_B}{\varepsilon r} +
	 \left(1-\frac{1}{\varepsilon} \right) \frac{2a_B}{R}
	 \sum_{l=1}^\infty \frac{l P_l(x)}{l+\varepsilon+l\varepsilon)} \cdot \frac{R^{2l}}{r_+^l r_-^l};
$$
$$
\frac{U_c(r_+>R,r_->R)}{\rm Ry} = \frac{2a_B}{\varepsilon r_+}
	 \left( 1+\sum_{l=1}^\infty \frac{(1+2l)P_l(x)}{1+l+l/\varepsilon} \cdot \frac{r_-^l}{r_+^l} \right).
$$
Here, the argument of the Legendre polynomials is $x \equiv \cos \vartheta$, where $\vartheta$ is the angle between the $z$ axis and the direction of ${\bf r}_-$. Note that the summation of these series is simplified considerably when using the following recurrent relationship
$$
 P_l(x) = [(2l-1)xP_{l-1}(x) -(l-1)P_{l-2}(x)]/l.
$$
Particular dependencies of $U_c$ for some selected arrangements of e$^+$ and e$^-$ and the cavity are shown in Fig. \ref{BaT_Uc}. Thus, we are able to take into account the variation of the e$^+$e$^-$ Coulombic energy during the formation of the Ps bubble. Similarly, the dielectric screening is used in the polaron problem and the ion-electron recombination problem (Onsager's formula) \cite{Pek51,Knox63}.

\subsection{Wave function of the e$^+$e$^-$ pair and minimization of its total energy $\langle H \rangle$}

\begin{figure}[t!]
\centering
 \includegraphics[width=120mm]{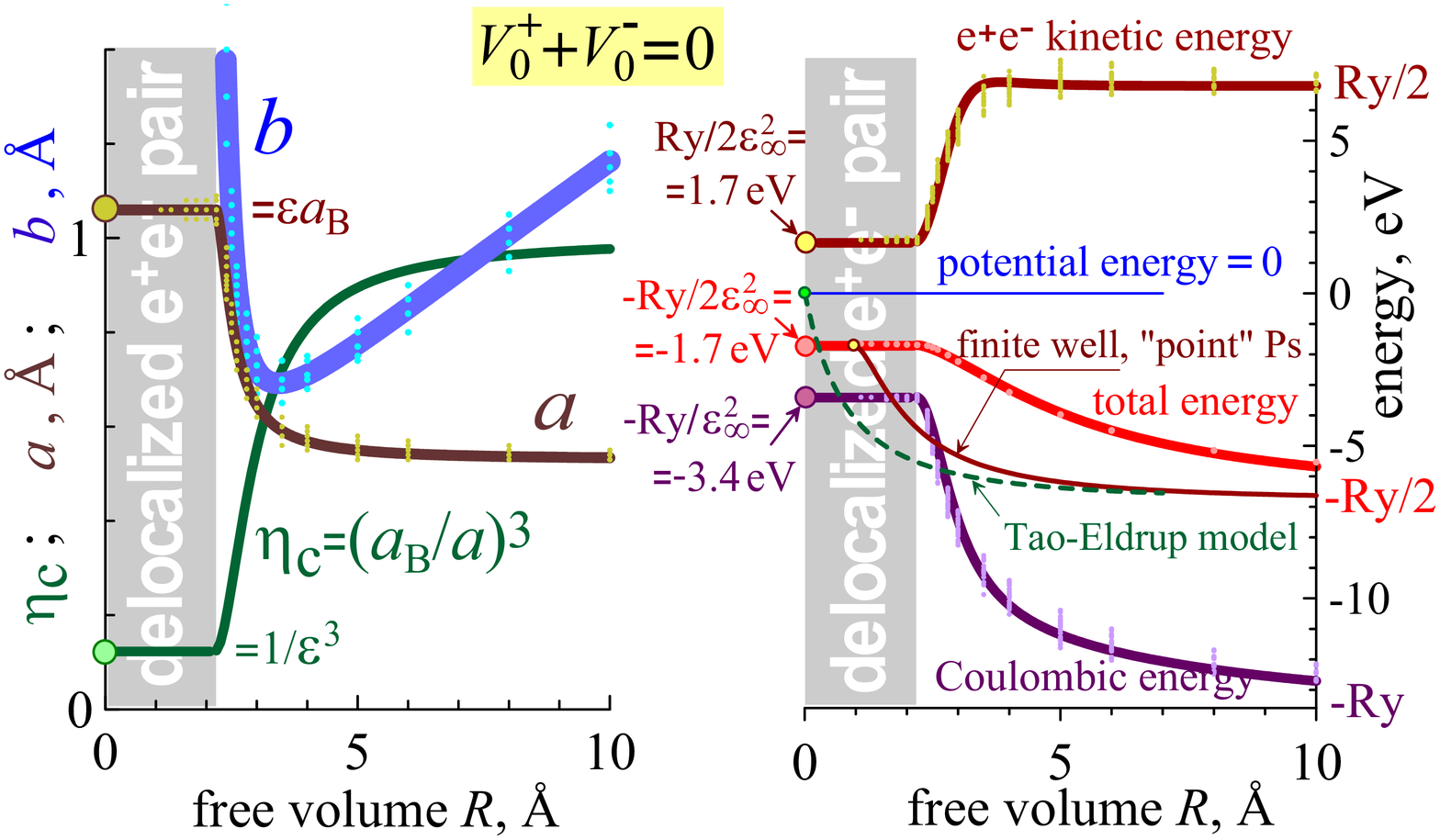}
\bigskip
 \includegraphics[width=120mm]{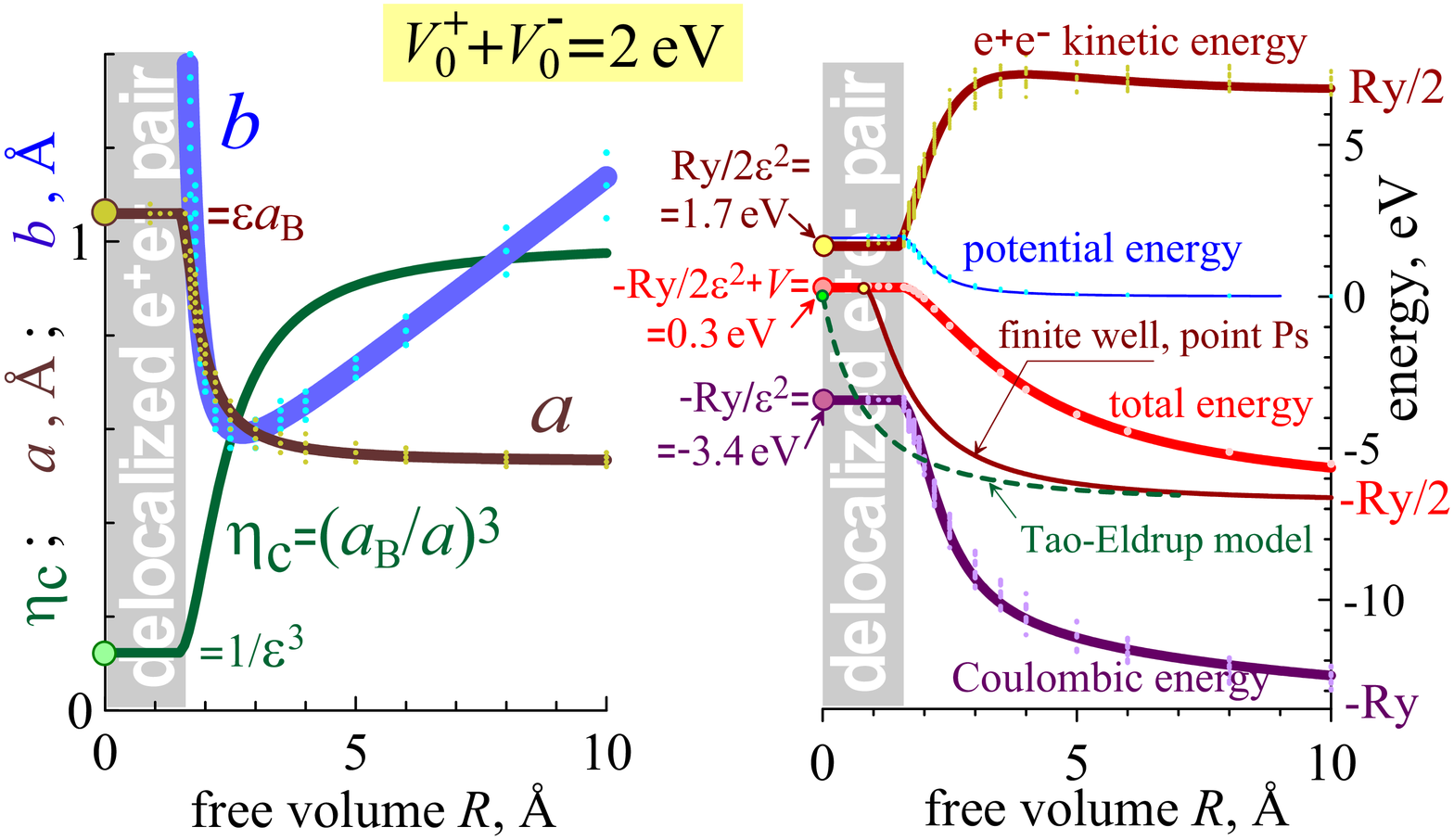}
\caption{Dependencies of the optimal parameters $a$ and $b$ vs. $R$, the bubble radius. They enter the e$^+$e$^-$ wave function and yield the minimum of the total energy $\langle H \rangle$. The relative contact density $\eta_c$ and different energy contributions to $\langle H \rangle$ (at optimal  $a$ and $b$) are shown as well. The upper drawings correspond to the case $V_0^+ + V_0^- =0$ and the lower ones to $V_0^+ + V_0^- =2$ eV. In both cases it was assumed that $\varepsilon = 2$.}\label{E_R}
\end{figure}

Keeping in mind further use of the variational procedure, let us choose the normalized e$^+$e$^-$ wave function in the following simplest form:
\begin{equation}\label{EM6}
\Psi_{+-}({\bf r}_+, {\bf r}_-) =\frac{\exp(-r/2a -r_{cm}/2b)}{8\pi\sqrt{a^3 b^3}},
 \quad
{\bf r}_{cm}=\frac{{\bf r}_+ + {\bf r}_-}{2},
 \quad
{\bf r}={\bf r}_+ - {\bf r}_-.
\end{equation}
In both cases of a rather large bubble and a uniform dielectric continuum, $\Psi_{+-}$ breaks into a product of two terms: the first one depends on the distance $r$ between e$^+$ and e$^-$, and the second one depends on the center-of-mass coordinate ${\bf r}_{cm}$.  Parameters $a$ and $b$ are the variational ones, over which we have minimized the energy of the e$^+$e$^-$ pair:
\begin{equation}\label{2_3}
E(a,b,R) = \langle \Psi_{+-}|H|\Psi_{+-} \rangle \to min ~~~
 \Rightarrow ~~~\hbox{$a(R)$, ~$b(R)$}.
\end{equation}
The simplest verification of the calculations is to recover two limiting cases. In case of large bubbles ($R\to \infty$), one should reproduce the ``vacuum'' state of the Ps atom: its total energy must tend to $\rm -Ry/2 = -6.8$ eV, the kinetic energy to $\rm +Ry/2$ and the Coulombic energy to $\rm -Ry$. In case of small bubbles ($R\to 0$), the delocalized qf-Ps state must be reproduced. The Schr\"odinger equation for qf-Ps has the same form as for the vacuum Ps, but with the substitution $e^2 \to e^2/\varepsilon$. Then the total qf-Ps energy tends to $V_0^+ +V_0^- \rm -Ry/2\varepsilon^2$, its kinetic part tends to $\rm +Ry/2\varepsilon^2=1.7$ eV ($\varepsilon = 2$) and the Coulombic energy tends to $\rm -Ry/\varepsilon^2=-3.4$ eV. Fig. \ref{E_R} displays optimal values of $a$ and $b$ as well as different contributions to the total energy of the e$^+$e$^-$ pair when $V_0^+ + V_0^- =0$ and 2 eV.

\subsection{Relative contact density and pick-off annihilation rate}

Using the wave function (\ref{EM6}) it is easy to obtain the relative contact density $\eta_c$ in the Ps atom:
\begin{equation}\label{B7_2}
\eta_c = \frac{\int \int d^3 {\bf r_+} d^3 {\bf r_-} |\Psi      _{+-}({\bf r}_+, {\bf r}_-)|^2 \delta({\bf r}_+ - {\bf r}_-)}
			  {\int \int d^3 {\bf r_+} d^3 {\bf r_-} |\Psi^{vac}_{+-}({\bf r}_+, {\bf r}_-)|^2 \delta({\bf r}_+ - {\bf r}_-)}
	   = \frac{a^3_B}{a^3(R)}.
\end{equation}
This quantity determines the observable Ps annihilation rate constant (including the case with applied permanent magnetic field). The resulting dependencies of $\eta_c$ are shown in Fig. \ref{E_R} (on the left). Because, for qf-Ps,  parameter $a$ is equal to $\varepsilon a_B$, for qf-Ps the value of $\eta_c$ should be $1/\varepsilon^3 =1/8$, which is well recovered in numerical calculations. When $R$ increases, $\eta_c$ approaches unity, because $a$ tends to its vacuum value $a_B$.

\begin{figure}[t!]
\centering
\begin{minipage}[h]{70mm}
 \includegraphics[width=70mm]{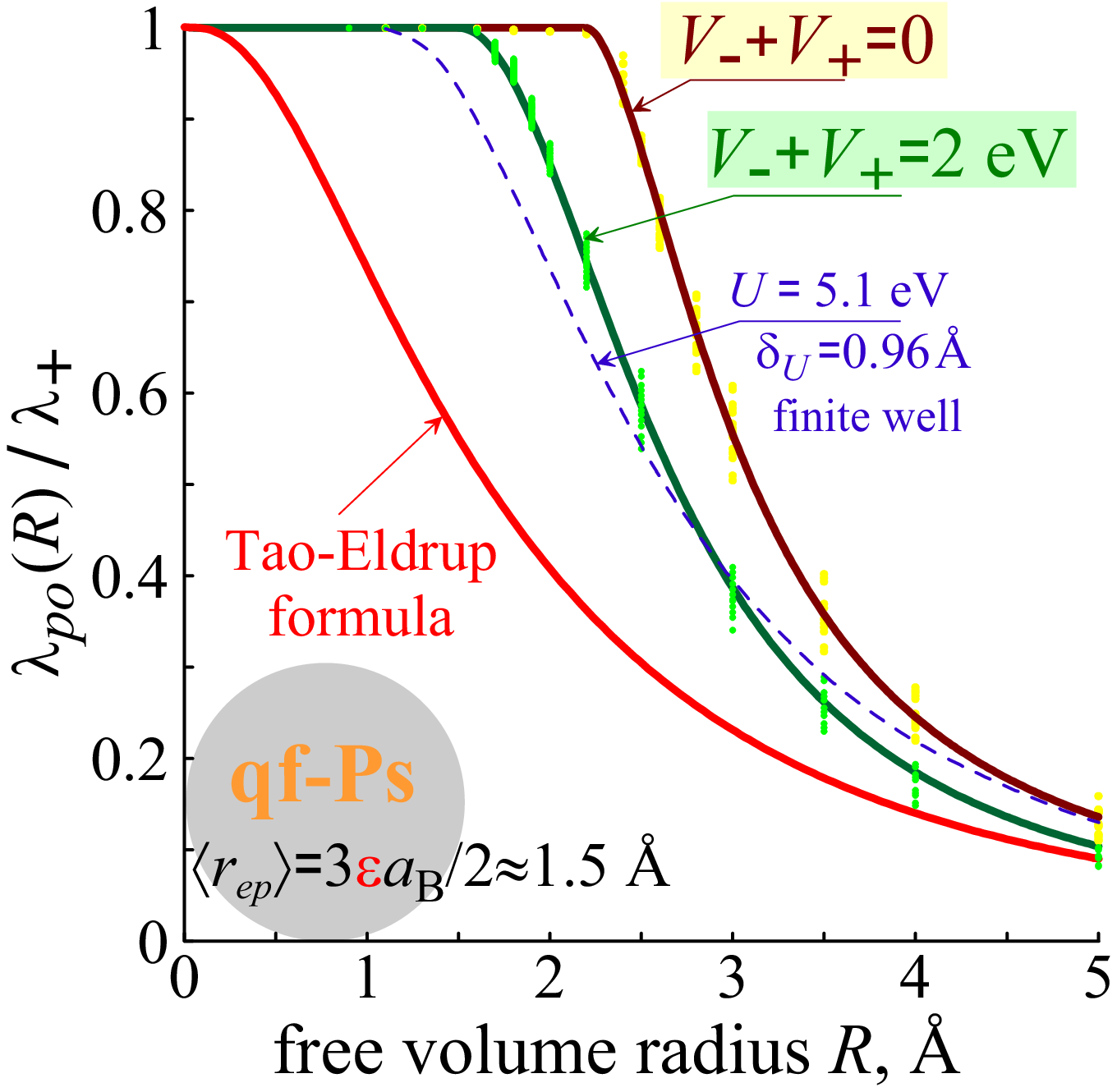}
\end{minipage}
\begin{minipage}[h]{80mm}
\caption{Pick-off annihilation rate constant of Ps, localized in a bubble of $R$ when
$V_+ + V_- =0$ (green curve) and $V_+ + V_- = 2$ eV (brown curve). For small $R$ ($\lesssim 2$ \AA)
the calculated values of $\lambda_{po}$ are equal to $\lambda_+$. The red line shows pick-off annihilation
rate constant, calculated according to the Tao-Eldrup formula. The dashed line is the calculation according to
the finite potential well model (for comparison we adopted that the depth of the well is
$(1-1/\varepsilon^2){\rm Ry}/2 \approx 5.1$ eV and its radius is $R$. The minimal radius of the well
when there appears an energy level is $\delta_U= 0.96$ \AA).}\label{L_po}
\end{minipage}
\end{figure}

Knowing the expression for the wave function (\ref{EM6}), one may calculate the positron overlapping $P_R$ with molecular electrons, surrounding the Ps atom, and therefore find out the pick-off annihilation rate constant: 
\begin{equation}\label{1_67a}
\lambda_{po}(R) \approx \lambda_+ P_R, \qquad
 P_R \approx \int_{r_+>R} d^3 {\bf r}_+ \int d^3 {\bf r}_- \left| \Psi_{+-}  ({\bf r_+} ,{\bf r_-}) \right|^2 .
\end{equation}
Here, $\lambda_+ \approx 2$ ns$^{-1}$ is the annihilation rate constant of ``free'' positrons. Results of calculations of $\lambda_{po}(R)$ for optimal $a$ and $b$ values, which correspond to the minimal Ps energy at a given $R$, are shown in Fig. \ref{L_po}.

\subsection{Discussion on the non-point-Ps approximation}

It is usually considered that Ps is a solvophobic particle, i.e., it forms a bubble when entering a liquid because of exchange repulsion between e$^-$ in Ps and the surrounding molecular electrons. If the work functions of e$^+$ and e$^-$ are negative ($V_0^+ \approx V_0^- < 0$), each particle considers a cavity as a potential barrier. So they are pulled to the bulk by polarization interaction with the medium. Nevertheless, even in this case the Ps bubble may be formed due to an enhancement of the Coulombic e$^+$e$^-$ attraction inside the cavity (no dielectric screening inside). This feature cannot be taken into account when Ps is simulated as a point particle.

It is seen that the behavior of the total energy of the pair (red curves in Fig. \ref{E_R}) strongly differs from the Tao-Eldrup prediction (green dashed curves; the first term in Eq. (\ref{9_2}), where $R_\infty$ is replaced by $R$), as well as from the expectation based on the finite potential well model (brown curves in Fig. \ref{E_R}; the Coulombic potential cannot be approximated well by a rectangular spherically symmetric potential). The same is true for the pick-off annihilation rate, Fig. \ref{L_po}.

Calculations demonstrate one common feature: up to $R\lesssim 1.5-2.2$ \AA\ all dependencies remain the same as in a medium without any cavity, but at larger $R$ there are significant deviations. This is related to the known quantum mechanical phenomenon -- absence of a bound state of a particle in a small finite 3d-potential well. In such cavities, Ps cannot be bound, it does not exert any repulsive pressure on their walls and does not stimulate their transformation towards the equilibrium Ps bubble. The possibility of finding a suitable preexisting cavity, sufficient at least for preliminary localization of qf-Ps, may be a limiting factor for the formation of the Ps bubble state.

One may find an equilibrium Ps bubble radius by minimizing the sum of the total e$^+$e$^-$ energy $\langle H \rangle$ and the surface energy of the bubble. For water it turns out to be 5-5.2 \AA\ which is about 2 \AA\ larger than predicted by the Tao-Eldrup model. For such a large bubble, the relative contact density is $\eta_c\approx 0.9$, Fig. \ref{E_R}. It is somewhat higher than the experimental values (0.65-0.75 \cite{11Ste_MSF}). This discrepancy may indicate that e$^+$ and e$^-$ really interact with a medium in a different way, for example, $V_0^- > V_0^+$. It means that the Ps electron may be trapped by a cavity, and e$^+$ will be bound to this trapped e$^-$ by the Coulombic attraction. This scenario may be also considered in the framework of the present approach, but the expression for the trial wave function of the new pair must be written in an ``asymmetric'' (towards e$^+$ and e$^-$) form:
\begin{equation}
\Psi_{+-}({\bf r}_+, {\bf r}_-) \approx \frac{\exp(-|{\bf r}_+ - {\bf r}_-|/2a  -r_-/2b)} {8\pi\sqrt{a^3 b^3}}.
\end{equation}

Any Ps bubble model reduces the original many-body (multi-particle) problem to a simpler one, that of one or two particles in an external field, which simulates the interaction with the medium. To calculate this field one usually relies on some macroscopic approaches. However, their validity always remains uncertain (for example, how to relate the actual arrangement of molecules around the Ps bubble with the jump of dielectric permittivity outside the bubble and so on).

\section{ Conclusion }

In this chapter an attempt has been made to trace the fate of a positron in molecular liquids, starting from its entering the medium till annihilation with an electron.

Energetic restrictions have been shown to leave only one possibility to form Ps, when the Ps atom is formed as a result of combination of the thermalized positron with one of the track electrons (some it is called as the (re)combination mechanism or the intratrack mechanism). This reaction proceeds in the e$^+$ blob, terminal part of the e$^+$ track. Considering the ionization slowing down of the projectile positron, one may conclude that the e$^+$ blob contains several tens of ion-electron pairs, which are confined in a spherical region with a radius of several nanometers. Ion-electron pairs in the blob are bound together because of their mutual Coulombic attraction. This is why further expansion (out-diffusion) of the blob is characterized as the ambipolar diffusion.

When the positron energy becomes less than $\approx$Ry=13.6 eV (subionizing e$^+$), this particle can escape from the blob. It becomes thermalized either outside the blob or inside it. Being inside the blob it may pick up intratrack electron and form Ps.

Primary ion-electron pairs transform into secondary radiolytic products (typical concentration $\gtrsim 0.01$ M), which are chemically active and thus can interact with Ps: oxidize it and/or stimulate its ortho-para conversion. Because the number of ion-electron pairs is large and the diffusion displacement of the species on a nanosecond timescale is comparable with the size of the blob (at least at a high-$T$ region), it appears inescapable to describe these processes on the basis of inhomogeneous diffusion kinetic equations in terms of concentrations of the species, rate constants, diffusion coefficients etc.

In the case of water it is clearly seen that taking into account interactions with intratrack products (Ps oxidation by OH radicals and $\rm H_3O^+$) is important. Otherwise, it is not possible to explain the experimentally observed decrease of the ortho-Ps lifetime with temperature.

Calculating the pick-off annihilation rate constant is of the utmost importance. This is the main goal of the Ps bubble models. Up to now most of calculations deal with the ``point-like''-positronium, by using its center-of-mass wave function only. In the framework of this approximation it is not possible to take into account the variation of the internal (Coulombic) energy of the e$^+$-e$^-$ pair during Ps bubble formation. However, we have shown that this energy contribution is not small and may play a decisive role.

Finally, we would like to draw attention to one of the possible applications of the positron spectroscopy: fast detection of potentially carcinogenic chemical compounds. These substances are considered as one of the major causes of human cancer. Modern technologies have led to new potential chemical carcinogens with which people may be in contact in everyday life. The European Community considers that about 100 such compounds need to be screened every day. Therefore, there is a need for a fast and cheap method of testing such compounds. Positron annihilation spectroscopy may provide such a method \cite{09Bya_MSF, 09Bya_PSSC}.

\medskip

This work is supported by the Russian Foundation of Basic Research (grant 11-03-01066) and the Russian Federal Agency of Atomic Energy.

\bigskip
\section*{Bibliography}
\bibliographystyle{unsrt}
\bibliography{./6_AMPC}

\end{document}